\documentclass[a4paper,11pt]{article}
\usepackage{bbm} 
\usepackage[dvips]{graphicx}
\usepackage{verbatim}
\usepackage{amsmath}
\usepackage{amsfonts}
\usepackage{amssymb}
\usepackage{latexsym}
\usepackage{color}
\usepackage{eufrak}
\usepackage{subfigure}
\newcommand{\be}[1]{\begin{equation} #1 \end{equation}}

\newcommand{\nzero}{N_0}
\newcommand\beq{\begin{equation}}
\newcommand\eeq{\end{equation}}
\newcommand{\bqa}{\begin{eqnarray}}
\newcommand{\eea}{\end{eqnarray}}  
\def\bra{\langle}
\def\ket{\rangle}

\def\Tr{\hbox{\rm Tr}}

\usepackage{ulem}

\makeatletter
\@addtoreset{equation}{section}

\makeatother

\begin{document}

\thispagestyle{empty}
\begin{flushright}
IFUP-TH/2007-16; SISSA-47/2007/EP\\
{\tt arXiv:0706.3854} \\
June, 2007 \\
\end{flushright}
\vspace{3mm}

\begin{center}

{\Large \bf
Non-Abelian vortices and monopoles  in $SO(N)$ theories}

\vskip 1.5em

\vskip 1.5em

{\large \bf  L.\,Ferretti$^{1,2}$}\footnote{\it  e-mail address:
ferretti@sissa.it}, \,\, 
{\large \bf   S.B.\,Gudnason $^{3,4}$} \footnote{\it  e-mail address:
gudnason@df.unipi.it},\,\, 
{\large \bf   K.\,Konishi$^{3,4}$} \footnote{\it  e-mail address:
konishi@df.unipi.it},

\vskip 1.5em

\vskip 1.5em

{\footnotesize 

\noindent $^1$ { SISSA,
via Beirut 2-4
I-34100 Trieste, Italy
}
\\
$^2$ { INFN, Sezione di Trieste,
I-34012 Trieste (Padriciano),  Italy }
\\
$^3$ { Department of Physics, ``E. Fermi'',  University of Pisa,
Largo Pontecorvo, 3,   Ed. C,  56127 Pisa, Italy }
\\[-2pt]
$^4$ { INFN, Sezione di Pisa,  
Largo Pontecorvo, 3, Ed. C, 56127 Pisa, Italy
}
}
\end{center}

\vspace{10mm}
%

\noindent {\large \bf Abstract:} 

  {Non-Abelian BPS vortex solutions are
  constructed  in   $\mathcal{N}=2$ theories with
  gauge groups  $SO(N)\times U(1)$.  The model has $N_{f}$ flavors of chiral multiplets in the vector representation of $SO(N)$, 
  and we consider a color-flavor locked vacuum in which the gauge symmetry is
   completely broken, leaving a global $SO(N)_{C+F}$ diagonal symmetry
  unbroken.  Individual vortices break this symmetry,  acquiring 
   continuous non-Abelian orientational moduli.  By embedding this model in high-energy theories with a hierarchical
  symmetry breaking pattern such as $SO(N+2)\rightarrow
  SO(N)\times U(1)\rightarrow  {\mathbbm {1}}$,  the correspondence between non-Abelian monopoles and vortices can be established through homotopy maps and flux matching, generalizing the known results in $SU(N)$ theories.  We find some interesting hints about the dual (non-Abelian) transformation properties among the monopoles.

  } 

\vfill

\newpage

\section{Introduction}

Recently some significant steps have been made in  understanding the non-Abelian monopoles \cite{NAmonop,GNO,BS,EW,Coleman,CDyons,DFHK,ABEKM}, occurring in spontaneously broken  gauge field theories \cite{Duality,Konishi}.  The basic observation is that the regular 't Hooft-Polyakov-like magnetic monopoles occurring in a 
system
\beq
 G   \,\,\,{\stackrel {v_{1}} {\longrightarrow}} \,\,\, H  \ ,
\label{system}   \eeq
where $H$ is a non-Abelian ``unbroken'' gauge group,  are not objects
which transform among themselves under the unbroken group $H$, but
which transform, if any,  under the {\it magnetic dual} of $H$,
namely ${\tilde H}$.  As field transformation groups, $H$ and ${\tilde H}$ are relatively non-local, thus a local transformation in the magnetic group ${\tilde H}$ would look like a non-local transformation in the electric theory.  Although this was implicit in the work by Goddard-Nuyts-Olive \cite{GNO} and others \cite{BS,EW},  the lack of the concrete knowledge on how ${\tilde H}$  acts on semiclassical monopoles has led to long-standing puzzles and apparent difficulties \cite{CDyons,DFHK}. 

Detailed study of gauge theories with $\mathcal{N}=1$ or $\mathcal{N}=2$ supersymmetry and quark multiplets,  on the other hand,  shows that light  monopoles  transforming as multiplets of non-Abelian magnetic gauge group ${\tilde H}$  do occur quite regularly  in full quantum systems \cite{APS,HO,CKM,BKM}.  They occur under certain conditions, e.g., that there is a
sufficiently large exact flavor symmetry group in the underlying
theory, which dresses the monopoles with flavor quantum numbers,
preventing them from interacting too strongly.   Also, the symmetry
requirement (i.e.~the symmetry of the low-energy effective theory describing the light monopoles be the correct symmetry of the underlying theory) seems to play an important role in determining the low-energy degrees of  freedom in each system \cite{MKY}.
There are subtle, but perfectly clear, logical reasons behind these quantum mechanical realizations of dual gauge symmetries 
in supersymmetric models.  Since there are free parameters in these supersymmetric theories which allow us to 
move from the fully dynamical regime to  semiclassical regions,
without qualitatively changing any physics, it must be possible to understand these light degrees of freedom in terms of more familiar soliton-like objects, e.g., semiclassical monopoles. 

This line of thought has led us to study the system (\ref{system}), in a regime of hierarchically broken gauge symmetries
 \beq
 G   \,\,\,{\stackrel {v_{1}} {\longrightarrow}} \,\,\, H  
 \,\,\,{\stackrel {v_{2}} {\longrightarrow}} \,\,\,
 {\mathbbm {1}} \ ,   \qquad  v_{1} \gg  v_{2} \ ,
\label{systembis}   \eeq
namely, in a phase in which the ``unbroken'' $H$ gauge system is completely broken at much lower energies (Higgs phase),  
so that one expects $-$ based on the standard electromagnetic duality
argument $-$ the ${\tilde H}$ system to be in confinement  phase.  The ``elementary monopoles''  confined by the confining strings in ${\tilde H}$ theory should look like 't Hooft-Polyakov monopoles embedded in a larger picture where their magnetic fluxes are frisked away by a magnetic vortex of the $H$ theory in Higgs phase.    

Indeed, in the context of softly broken $\mathcal{N}=2$  models,  this kind of systems can be realized concretely, by tuning certain  
free parameters in the models, typically, by taking the bare quark masses $m$ (which fix the adjoint scalar VEVs, 
$\bra \phi\ket = v_{1} \sim m$) much larger than the  bare adjoint scalar mass  $\mu$  (which sets the scale for the squark 
VEVs, $\bra  q  \ket = v_{2} \sim \sqrt{\mu m} $). In a high-energy approximation, where $v_{2}$ is negligible, 
one has a system, (\ref{system}), with a set of 't Hooft-Polyakov monopoles.  In the class of supersymmetric models considered, these monopoles are BPS, and their (semiclassical) properties are well understood.  In the low-energy approximation (where the massive monopoles are integrated out and $v_{1}$ is regarded as infinitely large)  one has the $H$ theory in Higgs phase, with BPS  vortices whose properties can also be studied in great detail. 

When the full theory is considered, with ``small'' corrections which involves factors of $\frac{v_{2}}{v_{1}}$,
there is an important qualitative change to be taken into account at the two sides of the mass scales (high-energy and low-energy). 
Neither monopoles of the high-energy approximation nor the vortices of the low-energy theory, are  BPS saturated any longer.  
They are no longer topologically stable.  This indeed follows from the fact that  $\pi_{2}(G)$ is trivial for any Lie group (no regular monopoles if $H$ is completely broken) or if  $\pi_{1}(G)={\mathbbm 1}$
(there cannot be vortices).  If   $\pi_{1}(G) \ne {\mathbbm 1}$ there may be some stable vortices left, but still there will be much fewer stable vortices as compared to what is expected in the low-energy theory (which ``sees'' only $\pi_{1}(H)$).  As the two effective theories  must be, in some sense, good approximations as long as   $\frac{v_{2}}{v_{1}} \ll 1$, one faces an apparent paradox.  

The resolution of this paradox is both natural and useful.  The
regular monopoles are actually sources (or sinks) of the vortices seen
as stable solitons in the low-energy theory; vice versa, the vortices
``which should not be there'' in the full theory, simply end at a
regular monopole.  They both disappear from the spectrum of the
respective effective theories.   This connection, however, establishes
one-to-one correspondence between a regular monopole solution of the
high-energy theory and the appropriate vortex of the low-energy
theory. As the vortex moduli and non-Abelian transformation properties
among the vortices, really depend on the exact global symmetry of the
full theory (and its breaking by the solitons),  such a correspondence
provides us with a precious hint about the nature of the non-Abelian
monopoles.   In other words, the idea  is to make use of the better
understood non-Abelian {\it vortices} to infer precise conclusions
about the non-Abelian {\it monopoles}, by-passing the difficulties
associated with the latter as mentioned earlier.  

A  quantitative formulation of these ideas requires a concrete knowledge of the vortex moduli space and the transformation properties among the vortices \cite{Hashimoto,Auzzi:2005gr,seven}. This problem has been largely clarified, 
 thanks to our generally improved understanding of non-Abelian vortices \cite{HT,ABEKY,SY,ABEK,Tong},
and in particular to the technique of the ``moduli matrix''
 \cite{Isozumi:2004vg},  especially  in the context of $SU(N)$ gauge theories.  Also,  some puzzles
 related to the systems with symmetry breaking $SO(2N) \to  U(N),$ or
 $SO(2N) \to U(r)\times   U(1)^{N-r}$, have found natural solutions   
\cite{Duality}.  

In this article, we wish to extend these analyses to the cases
  involving vortices of $SO(N)$ theories.  In \cite{FK} the first
  attempts have been made in this direction, where  softly broken
  $\mathcal{N}=2$ models with $SO(N)$ gauge groups and with a set of
  quark matter in the vector representation,    have been analyzed.
  In the case of $SO(2N+3)$ theory broken to $SO(2N+1) \times U(1)$
  (with the latter completely broken at lower energies) one observes
  some hints how the dual, $USp(2N)$ group, might emerge.  
  In the model considered in \cite{FK}, however, the construction of  the system in which the gauge symmetry is completely broken, leaving a maximum exact color-flavor symmetry  (the color-flavor locking),   required  an ad hoc addition of an $\mathcal{N}=1$ superpotential, in contrast to $SU(N)$ theories where, due to the vacuum alignment with bare quark masses familiar from $\mathcal{N}=1$ SQCD, the color-flavor locked vacuum  appears quite automatically.  

In this article we therefore  turn to a slightly different class of  $SO(N)$  models.  
The underlying theory is an $SO(N+2)$  gauge theory with matter hypermultiplets in the adjoint representation, with the gauge group broken partially  at a mass scale $v_{1}$.    
   The analysis is
slightly more complicated than the models considered in \cite{FK}, but in the present model  the color-flavor locked vacua  occur  naturally.  
Also, these models have a richer spectrum of vortices and monopoles
than in the case of \cite{FK}, providing us with a finer testing
ground for duality and confinement. 

At scales much lower than $v_1$, the model reduces to an $SO(N)\times U(1)$ theory  with quarks in the {\it vector} representation.  Non-Abelian vortices arising in the color-flavor locked vacuum of this theory  transform   non-trivially under the  $SO(N)_{C+F}$ symmetry. We are interested in their role in the dynamics of gauge theories, but these solitons also play a role in cosmology and condensed matter physics, so the results of sections \ref{sec:2Nvor}  and \ref{Modd} of this paper could be  of more general interest (for example they can be useful for cosmic strings, see  \cite{Eto:2006db}).

In section \ref{sec:model} of this article, we present the high-energy model with gauge group $SO(2N+2)$ . In section \ref{sec:2Nvor} we study its low-energy effective theory and present the vortex solutions. In section \ref{Modd} we study the model with gauge group $SO(2N+3)$. Finally, in section \ref{corres} we discuss the correspondence between monopoles and vortices.


\section{The model  \label{sec:model}}

We shall first discuss the $SO(2N+2)$  theory; the case of $SO(2N+3)$  group will be considered separately later. 
We wish to study the properties of monopoles and vortices occurring in the system
\begin{equation} SO(2N+2) \stackrel{v_1}{\longrightarrow} SO(2N)
  \times U(1) \stackrel{v_2}{\longrightarrow} {\mathbbm {1}} \ .    \label{hierarchy} \end{equation} 
To study the consequences of such a breaking, we take a concrete
example of an $\mathcal{N}=2$ supersymmetric theory with gauge group
$SO(2N+2)$ and $N_f$ matter hypermultiplets in the adjoint
representation. All the matter fields have a common mass $m$, so the
theory has a global $U(N_f)$ flavor symmetry.   We also add a small
superpotential term $\mu \phi^2$ in the Lagrangian, which breaks softly $\mathcal{N}=2$ to $\mathcal{N}=1$.   For the purpose of considering hierarchical symmetry breaking (\ref{hierarchy}), we take 
\beq   m  \gg    \mu \ .
\eeq
The theory is infrared-free for $N_{f} > 1$, but one may consider it
as an effective low-energy theory of some underlying theory, valid at
mass scales below a given ultraviolet cutoff.  In any case, our
analysis will focus on the questions how the properties of the
semiclassical monopoles arising from the intermediate-scale   can be
understood through the  moduli of the non-Abelian vortices arising
when the low-energy, $SO(2N)$ theory is put in the Higgs phase.

The superpotential of the  theory has the form, 
\beq  W=  \sqrt{2}  \sum_{A}   \Tr \, {\tilde \zeta}_{A} \left[ \phi, \zeta_A
  \right]+  m   \sum_{A}  \Tr \,  {\tilde \zeta}_{A}  \zeta_A   +  
\frac{\mu}{2}  \Tr \, \phi^{2}  \ . 
\label{superp}  \eeq
In order to minimize the misunderstanding, we use here the notation of $\zeta_A$, $ {\tilde \zeta}_{A}$
for the quark hypermultiplets in the {\it adjoint} representation of the high-energy 
gauge group $SO(2N+2)$  (or $SO(2N+3)$), with $A=1,2,\ldots, N_{f} $ standing for the flavor index.  We shall reserve the symbols $q_{A}, {\tilde q}_{A}$ for the light supermultiplets of the low-energy theory, which transform as the {\it vector} representation of the gauge group $SO(2N)$  (or $SO(2N+1)$).
The vacuum equations for this theory therefore take the form
\begin{align}
\left[\phi, \phi^\dagger\right]&=0 \ , \label{ve1} \\
\sum_A \left[\zeta_A,{\zeta}^\dagger_A\right]&=\sum_A
  \left[{\tilde{\zeta}}^\dagger_A,\tilde{\zeta}_A\right] \ ,  \label{ve2} \\
\sum_A \sqrt{2}\left[\zeta_A,\tilde{\zeta}_A\right]+\mu\, \phi&=0 \ , \label{ve3} \\
\sqrt{2}\left[\phi,\zeta_A\right]+m\, \zeta_A&=0 \ , \label{ve4} \\
-\sqrt{2}\left[\phi,\tilde{\zeta}_A\right]+m\, \tilde{\zeta}_A&=0 \ . \label{ve5}
\end{align}
We shall choose a vacuum in which  $\phi$ takes the vacuum expectation
value  (VEV)
\begin{equation}
\langle\phi\rangle=\begin{pmatrix}
 0 & -iv & 0& \cdots&0  \\
 iv & 0 &0 &\cdots &0  \\
0 & 0& 0 &\cdots & 0 \\
 \vdots&\vdots & \vdots &\ddots  & 0\\
 0& 0& 0& 0&  0\\
\end{pmatrix} \ ,
\end{equation}
 which breaks $SO(2N+2)$ to $SO(2N)\times U(1)$ and is consistent with Eq.~(\ref{ve1}).

We are interested in the Higgs phase of the theory. 	In order for the $SO(2N)\times U(1)$ symmetry to be broken at energies much lower than  $v_{1}\equiv v$, we have to find non-vanishing VEVs of the squarks which satisfy Eqs.~(\ref{ve4}),(\ref{ve5}). This means that $v \sim \mathcal{O}(m)$.
The magnitude of squark VEVs is then fixed by Eq.~(\ref{ve3})  to be of the order of  $(\mu\, m)^{1/2}  \ll   m$ and defining $v_2\equiv  |\bra q \ket | = \mathcal{O}(\sqrt{\mu m})$ we obtain the hierarchical breaking of the gauge group (\ref{hierarchy}). The D-term condition (\ref{ve3}) can be satisfied by the ansatz
 \be{  \zeta =   {\tilde \zeta}^{\dagger} \ .}
One must also determine the  components of the fields $\zeta, {\tilde \zeta}$  which do not get a mass of the order of  $\mathcal{O}(v)\simeq\mathcal{O}(m)$. We see from Eq.~(\ref{superp})  that the light squarks are precisely those for which Eqs.~(\ref{ve4}),(\ref{ve5})  are satisfied non-trivially, 
i.e.,  by non-vanishing  ``eigenvectors''  $\zeta$, ${\tilde \zeta}$. The conditions (\ref{ve4}),(\ref{ve5}) require that the light components correspond to the generators of $SO(2N)$ which are lowering and raising operators for $\langle{\varphi}\rangle$. This condition implies also  
 \be{ v=\frac{m}{\sqrt{2}} \ .  }

To find the light components of $\zeta, {\tilde \zeta}$,  we note that for a
single flavor,
Eqs.~(\ref{ve3})-(\ref{ve5}) together have the form of an
${\EuFrak{su}}(2)$ or ${\EuFrak{so}}(3)$ algebra,  $T_{1},T_{2},T_{3}$,  
\beq      \phi \propto  T_{3} \ , \qquad    \zeta_{A}  \propto   T_{-}=
 T_{1}- i T_{2} \ , \qquad    {\tilde \zeta}_{A} \propto   T_{+}= T_{1} +
 i T_{2} \ , 
\eeq 
with appropriate constants.

  The simplest  way to proceed  is to consider the various  $SO(3)$ subgroups,  $SO(3)_{12j}$,   lying in  the $(1 2 j )$ three-dimensional subspaces  ($j=3,4,5,\ldots$),    with
\beq    T_{3}=  H^{(0)}=  -i  \Sigma_{12} =
\left(\begin{array}{ccc}   0 & -i & 0   \\ i & 0 & 0  \\ 0 & 0 & 0
\end{array}\right)_{12j} \ , 
\eeq
\beq   T_{-} = T_{1}- i T_{2} =  L_{j, -} \equiv  \left(\begin{array}{ccc}  0 & 0 &  1  \\  0 & 0 &  -i  \\  -1 &  i &  0  
\end{array}\right)_{12j}, \qquad   T_{+}=  T_{1}+ i T_{2} = L_{j,
  -}^{\dagger} \ .    \label{simpleway}
\eeq
  The light fields which remain massless can then be expanded as 
\be{\zeta_A(x)  = \sum_{j=3,4, 5 ,\ldots}  \frac{1}{2}\,   q_{jA}(x) \,
  L_{j, -}  \ ,  \qquad   {\tilde \zeta}_{A}(x) =  \sum_{j=3,4, 5 ,\ldots}  \frac{1}{2}\,   \tilde{q}_{Aj}(x) \,
  L_{j, +} \label{Expanded}}  
for each flavor  $A=1,2, \ldots,  N_{f}$.   Written as a full $SO(2N)$ matrix, $ L_{j, -} $  looks like 
\beq 
 L_{j, -}=\left(\begin{array}{ccccc}0 & 0 & \ldots &  1 & \ldots \\0 &
   0 &  & - i & \vdots \\\vdots &  & \ddots &  &  \\  -1 & i &  &  &
   \vdots \\\vdots & \ldots &  & \ldots & 0\end{array}\right) \ ,
   \qquad  L_{j, +}=  L_{j, -}^{\dagger}  \ .
 \eeq
In $L_{j, -}$  the only non-zero elements  ($1$ and $-i$)  in the first two rows  appear in the $ (2+j)$-th  column; the only two non-zero elements 
 in the first two columns ($-1$ and $i$)  appear in the $(2+j)$-th  row.  

 An alternative way to find the combinations which do not get mass from $\bra \phi \ket$   is to use the  independent $SU(2)$ subgroups contained in various
$SO(4)$ subgroups  living in the subspaces  $(1,2, j, j+1)$,   $
 j=3,5,\ldots,  2N-1$. As is
 well known, the ${\EuFrak{so}}(4)$ algebra factorizes into two {\it
   commuting} ${\EuFrak{su}}(2)$ algebras,  
\beq    {\EuFrak{so}}(4) \sim {\EuFrak{su}}(2)\times {\widehat
  {\EuFrak{su}}(2)} \ ,  
\eeq  
where for instance for  $SO(4)_{1234}$ one has  
\beq
S_1 = -\frac{i }{ 2} ( \Sigma_{23} +  \Sigma_{41} )\ , \quad
S_2 = -\frac{i }{ 2} ( \Sigma_{31} + \Sigma_{42} )\ , \quad
S_3 = -\frac{i }{ 2} ( \Sigma_{12} +  \Sigma_{43} )\ ,  \label{su21}
\eeq
\beq
{\hat S}_1 = -\frac{i }{ 2} ( \Sigma_{23} -  \Sigma_{41} )\ , \quad
{\hat S}_2 = -\frac{i }{ 2} ( \Sigma_{31} -  \Sigma_{42} )\ , \quad
{\hat S}_3 = -\frac{i }{ 2} ( \Sigma_{12} -  \Sigma_{43} )\ , \label{su22}
\eeq
where  
\[ \Sigma_{23} =  \left(\begin{array}{cc}0 & 1 \\-1 & 0\end{array}\right)_{23},  \] is (up to a phase) the  rotation generator  in the $23$ plane, etc.

Since 
 \beq \frac{\sqrt{2}}{m} \,\bra \phi \ket  =H^{(0)} =  -i
 \,\Sigma_{12}  = S_3 + {\hat S}_3 \ ,    \label{cfr}\eeq  
it follows from the standard  ${\EuFrak{su}}(2)$ algebra  
that {\it both}\, $S_{-}=S_{1}-i S_{2}$ and ${\hat S}_{-}={\hat S}_{1}- i {\hat S}_{2}$    satisfy   the relation, 
\beq \left[ \frac{\sqrt{2}}{m}\bra \phi \ket,   S_{-} \right]=  -S_{-}\ ,\qquad
\left[ \frac{\sqrt{2}}{m}\bra \phi \ket,  {\hat S}_{-} \right]=  -{\hat S}_{-}\ .  \label{condlight}
\eeq
One can choose the two combinations 
 \beq L_{-}=   S_{-}  + {\hat S}_{-}\ ; \qquad L_{-}^{\prime}=  S_{-}
 -  {\hat S}_{-}\ , \label{twocomb}\eeq 
which satisfy the required relation,
\beq \left[\frac{\sqrt{2}}{m} \bra \phi \ket,   L_{-} \right]=  - L_{-}\
 ,\qquad  \left[\frac{\sqrt{2}}{m}\bra \phi \ket,   L_{-}^{\prime} \right]=  -
 L_{-}^{\prime}\ .   \label{condlightbis}
\eeq
These constructions can be done  in  all ${\EuFrak{su}}(2)$ subalgebras  living in $SO(4)_{1,2, j, j+1}$,  $j=3,5,\ldots, 2N-1$.

 Explicitly,  
 $ S_{j \, -}$,  ${\hat S}_{j\, -}$, 
     and   $L_{j, -}$,  $L_{j, -}^{\prime}$  have the form  ($j=3,5,\ldots$)
\beq{
 S_{j \, -}=\frac{1}{2} \left(\begin{array}{cccc}0 & 0 &  1 & i \\0 & 0 & -i & 1   \\  -1 &  i & 0 & 0  \\  
-i  & -1 & 0  & 0    
\end{array}\right)_{(1,2,j, j+1)}, \qquad  {\hat S}_{j\, -}= \frac{1}{2} \left(\begin{array}{cccc} 0 & 0 &  1 & -i  \\0 & 0 &  -i & -1   \\  -1  &  i  & 0  &0  \\  
i  & 1 & 0 & 0  \end{array}\right)_{(1,2,j, j+1)}};  \eeq
\beq {
L_{j, -}= \left(\begin{array}{cccc}0 & 0 &  1 & 0  \\  0 & 0 & -i  &  0\\  -1 &  i & 0 & 0 \\
0 & 0 &  0  &0  
\end{array}\right)_{(1,2,j, j+1)}, \qquad  L_{i, -}^{\prime} = \left(\begin{array}{cccc} 0 & 0 & 0 & i  \\0 & 0 &   0 & 1  \\   0 &  0  &  0  &0   \\  
-i  & -1 &  0 & 0  \end{array}\right)_{(1,2,j, j+1)}}. \label{Lequat} \eeq
Clearly, one can write 
\beq    L_{j, -}^{\prime}  = i \,  L_{j+1, -} \ ;
\eeq
and use the first of  Eq.~(\ref{Lequat}) to define $L_{j, -}$ for all $j=3,4,5,\ldots$,  $j$  even or odd.   With this definition,  $L_{j, -}$ coincide with those introduced in  Eq.~(\ref{simpleway})  by using various $SO(3)$ subgroups.

Eqs.~(\ref{superp}),(\ref{condlight}),(\ref{condlightbis}) show that the light fields (those which do not get mass of  order  $m$)  are the ones appearing in the expansion (\ref{Expanded}).   
Alternatively,  the basis  of light fields can be taken as 
 \be{\zeta_A(x)  =\frac{1}{\sqrt{2}}\, \sum_{i=3,5,\ldots} \left[\, Q_{iA}(x)
 \, S_{i, -}   +  {\hat Q}_{iA}(x)  {\hat S}_{i, -}\,\right]\ ,   \qquad
 {\tilde \zeta}_{A} =\frac{1}{\sqrt{2}}\, \sum_{i=3,5,\ldots} \left[\, \tilde{Q}_{Ai}(x)
 \, S_{i, +}   +  {\hat{\tilde{Q}}}_{Ai}(x)  {\hat S}_{i, +}\,\right]  \ .  \label{Expanbis}}
 The relation between the $q_{iA}(x)$  and  $Q_{iA}(x)$ fields is  ($i=3,5,\ldots $): 
 \beq   Q_{iA}(x)  = \frac {q_{iA}(x)  +  i \,q_{i+1,A}(x)}{\sqrt{2} }\ ; \qquad 
 {\hat Q}_{iA}(x)    =  \frac {q_{A,
 i}(x)   -  i\, q_{A, i+1}(x)}{\sqrt{2} }=  Q_{i+1,A}(x) \ .
\label{U2Ntra} \eeq
 All other components
get a mass of order $m$.    There are thus precisely  $2N$  light   quark fields  (color components) 
 $q_{iA}$  ($i=1,2,\ldots, 2N$) for each flavor.      These are the light hypermultiplets of the theory. 
 
 Each of the two bases $\{q_{iA} \}$  or  $\{ Q_{iA} \}$ has  some advantages.  
  Clearly the basis $q_{iA}$ ($i=1,2,\ldots, 2N$)  corresponds to the usual basis of the  fundamental (vector) representation of  the $SO(M)$ group  ($M=2N$), appearing in the decomposition of an adjoint representation of $SO(M+2)$ into the irreps of $SO(M)$:  
 \beq    \frac{(M+2)(M+1)}{2}  =    \frac{ M(M-1)}{2}    \oplus  M
  \oplus M   \oplus    1 \ .
 \eeq
The low-energy effective Lagrangian  can be most easily written down in terms of these fields,  and the symmetry property of the vacuum  is 
manifest  here.

On the other hand,  the basis $ (Q_{jA},  {\hat Q}_{jA}) $, $j=3,5,7,\ldots$,   is made of  pairs of eigenstates of the ($a\equiv  (j-1)/2$)-th  Cartan subalgebra generator, 
\beq      H^{(a)}  =   - i \, \Sigma_{j, j+1} = S_{j, 3} - {\hat
  S}_{j, 3}\  , \qquad a=\frac{j-1}{2} =  1,2,\ldots, N \ , 
\eeq
 (see Eqs.~(\ref{su21}),(\ref{su22}),(\ref{cfr})), with eigenvalues
$\pm 1$, so that the vortex equations can  be  better formulated,
and the symmetry maintained by individual  vortex solutions can be seen explicitly in this basis.  $Q_{iA},$  $(i=3,5,\ldots),$ form
an  $\underline{\mathbf{N}}$ of $SU(N) \subset SO(2N)$;   $ {\hat Q}_{iA},$  $(i=3,5,\ldots),$   form an  $\underline{\bar{\mathbf{N}} }$. In other words,  it represents the decomposition of a $\underline{\mathbf{2N} }$ of $SO(2N)$ into   $\underline{\mathbf{N} } + \underline{\bar{\mathbf{N}}} $ of $SU(N) \subset SO(2N)$. 
The change of basis from the 
vector basis ($q$) and  $U(N)$  basis  ($Q, {\hat Q}$)  is discussed  more extensively in Appendix A.

%


\section{Vortices in the $SO(2N)\times U(1)$  theory \label{sec:2Nvor}}

\subsection{The vacuum and BPS vortices\label{subsec:vacBPS}}
The low-energy Lagrangian for the theory with gauge group $SO(2N)\times U(1)$ and squarks $q_A$,$\tilde{q}_A$ in the fundamental representation of $SO(2N)$ is 
\begin{eqnarray}
\mathcal{L}&=&-\frac{1}{4g_{1}^2}F^{0\mu \nu}F^{0}_{\mu \nu}-\frac{1}{4g_{2N}^2}F^{b\mu \nu}F^{b}_{\mu \nu}
+ \left|\mathcal{D}_\mu q_{A}\right|^2+
 \left|\mathcal{D}_\mu\tilde{q}_{A}^\dagger\right|^2\\
 && - \frac{g_{2N}^2}{2}\left| q_A^\dagger t^{b}q_A-\tilde{q}_At^{b}\tilde{q}_A^{\dagger}\right|^2-2g_{2N}^2 
 \left|\tilde{q}_At^bq_A\right|^2 \nonumber \\
 && -\frac{g_1^2}{2}\left| q_A^\dagger q_A-\tilde{q}_A \tilde{q}_A^{\dagger}\right|^2 
  - 2 g_1^2 \left|\tilde{q}_A q_A  + \frac{\mu m}{\sqrt{2}}  \right|^2 + \cdots \nonumber
 \end{eqnarray}
where the dots denote higher orders in $\mu/m$ and terms involving
$\delta\phi=\phi-\langle\phi\rangle$. 
Note that to this order, 
the only modification is a Fayet-Iliopoulos term which does not break $\mathcal{N}=2$ SUSY. 
The covariant derivative acts as
\be {\mathcal{D}_\mu q_{A}=\partial_\mu q_A   -   i A_\mu^{0}\,q_A 
- i A^b_\mu t^b q_A \ ,}
where $t^{a}$ is normalized as
\beq  \Tr \, \left(t^{a}\right)^{2} =1 \ , \eeq
and
\beq \quad t^{a}= \frac{1}{\sqrt{2}}\,  H^{(a)} =
\frac{1}{\sqrt{2}}\,  \left(\begin{array}{cc}0 & -i \\i &
  0\end{array}\right)_{2a+1, 2a+2},
\label{sqrt2}\eeq
where $H^{(a)}$ is the $a$-th Cartan generator of $SO(2N)$,  $a=1,2,\ldots, N$,
 which we take simply as 
 \beq H^{(a)} = \left(\begin{array}{cc}0 & -i \\i & 0\end{array}\right)_{2a+1, 2a+2}.   \label{cartan}
\eeq
As we have seen already, each light field carries unit charge with respect to  $H^{(0)}$; 
  the pair  $ (Q_{A, j},  {\hat Q}_{A, j}) $, $j=3,5,7,\ldots$, furthermore  
carries the charge $\pm 1$ with respect to $ H^{(a)} $  ($a= (j-1)/2)$ and zero charge with respect to other Cartan generators.

Let us define 
\beq \xi=\frac{\mu  \, m}{2} \ , \eeq
which is the only relevant dimensional parameter in the Lagrangian. We set $N_f=2N$, which is enough for our purposes\footnote{Higher $N_f$ are interesting because of semilocal vortex configurations arising in these theories. These solutions will be discussed elsewhere.}.
By writing $q_{iA}$,  ${\tilde q}_{Ai}$   as  color-flavor mixed matrices $q$, ${\tilde q}$,  the vacuum equations are now cast into the form 
\begin{align}
\mathrm{Tr}\big(qq^\dagger\big)&=\mathrm{Tr}\big(\tilde{q}^\dagger\tilde{q}\big) \ , \\
qq^\dagger-\big(qq^\dagger\big)^T&=\tilde{q}^\dagger\tilde{q}-\big(\tilde{q}^\dagger\tilde{q}\big)^T
\ , \\
\mathrm{Tr}\big(q\tilde{q}\big) &=\xi \ ,  \\
\mathrm{Tr}\,  \big(t^{b} q {\tilde q\big)}&=0 \ .
\end{align}
The vacuum we choose to study is characterized by the color-flavor locked phase
\beq  \langle q_{A, j}\rangle=  \left\langle {\tilde{q}}_{A, j}^\dagger
\right\rangle= \delta_{A, j} \, v_{2} \ , \qquad   v_{2}  =  \sqrt{\frac{\xi}{2N}}\ ,   \eeq
or
\beq  \langle q\rangle=\big\langle\tilde{q}^\dagger\big\rangle=v_{2}\,
      {\mathbbm {1}} = v_{2}\,\left(\begin{array}{cccc}1 & 0 & 0 & 0
	\\0 & 1 & 0 & 0 \\0 & 0 & \ddots & 0 \\0 & 0 & 0 & 1
      \end{array}\right) \ ,   \label{squarkVev}\eeq
which clearly satisfies all the equations above. 
The gauge ($O$)   and flavor ($U$)  transformations act on them  as  
\be{q\rightarrow O\, q\, U^T \quad, \quad \tilde{q} \rightarrow U^* \,
  \tilde{q}\, O^T\qquad  O \in SO(2N)\times U(1) \ , \quad U \in U(2N):}  
 the gauge group is completely broken, while a global $SO(2N)_{C+F}\times U(1)_{C+F}$ group ($U=O$) is left   unbroken.

When looking for vortex solutions, one  suppresses time and $z$ dependence of the fields and retains only the component $F_{xy}$ of the field strength. The vortex tension can be cast in the Bogomol'nyi form
\begin{align}
T =\int d^2x \ \bigg\{&\left| \frac{1}{2\, g_{2N}} \,F^b_{ij}\pm g_{2N}\varepsilon_{ij}\tilde{q}_At^bq_A\right|^2 + 
 \left| \frac{1}{2\, g_1}F^{0}_{ij} \pm g_1\varepsilon_{ij}\left(\tilde{q}_Aq_A-  \xi\right)\right|^2
\nonumber \\ 
&+\frac{1}{2}\left| \mathcal{D}_iq_A \pm i \varepsilon_{ij}  \mathcal{D}_j \tilde{q}_A^\dagger\right|^2 + \frac{1}{2}\left| \mathcal{D}_i\tilde{q}_A^\dagger \pm  i \varepsilon_{ij}  \mathcal{D}_j q_A\right|^2 \nonumber\\
&+\frac{g_{2N}^2}{2}\left| q_A^\dagger t^{b}q_A-\tilde{q}_At^{b}\tilde{q}_A^{\dagger}\right|^2+\frac{g_1^2}{2}\left| q_A^\dagger q_A-\tilde{q}_A\tilde{q}_A^{\dagger}\right|^2 
   \pm   \varepsilon_{ij} \,  \xi  F^{0}_{ij} \bigg\} \ .  \label{tc}
   \end{align}
The terms with the square brackets in the last line of Eq.~(\ref{tc})  automatically vanish  with the ansatz \cite{ABEKY} 
\be{q_{iA}=\tilde{q}_{iA}^\dagger: \label{ans}} 
thus we shall use this ansatz for the vortex configurations.  The resulting 
BPS equations are
\begin{align}
\frac{1}{2\, g_1} F^{0}_{ij}+\eta \, g_1 \, \varepsilon_{ij} \, \left(q_A^\dagger
 q_A-\xi\right)& =0 \ , \label{bps1}\\
 \frac{1}{2\, g_{2N}}  \, F^b_{ij}+\eta \, g_{2N} \,\varepsilon_{ij} \, q_A^\dagger
 t^bq_A &=0 \ , \label{bps2}\\
 \mathcal{D}_iq_A+   i  \, \eta \,  \varepsilon_{ij} \,  \mathcal{D}_j q_A& =0\
 \label{bps3}, \qquad  \eta =\pm 1\ ,
\end{align}
 where we have used the ansatz (\ref{ans}). The tension for a BPS solution is
\beq 
T =  \eta \int d^2x  \, \varepsilon_{ij}\,  \xi \, 
F^{0}_{ij} \ .  \eeq
To obtain a solution of these equations, we need an ansatz for the squark fields. It is convenient to perform a $U(2N)_F$ transformation (\ref{U2Ntra}),  where  the vacuum takes the block-diagonal form
\begin{equation}
\langle  Q  \rangle=\langle\tilde{Q}^\dagger\rangle=\sqrt\frac{\xi}{2  N} \cdot \frac{1}{\sqrt{2}}\begin{pmatrix}
    1 & 1& 0  & 0& \cdots\\
    i &-i& 0  & 0& \cdots  \\
    0  & 0&  1&1& \cdots \\
    0& 0&i& -i&  \cdots  \\     
    \vdots& \vdots& \vdots& \vdots& \ddots   
\end{pmatrix} \ , 
\end{equation}
In this basis, the ansatz is:
\be{
A_i=h_a(r)\, t^a\, \varepsilon_{ij} \, \frac{r_j}{r^2}\ ; \qquad   t^{0}  \equiv \frac{1}{\sqrt{2}}\ , \quad t^{a} = \frac{1}{\sqrt{2}}  \left(\begin{array}{cc}0 & -i \\i &
  0\end{array}\right)_{2a+1, 2a+2}\ ;    \label{aaa}}
\be{
Q(r, \vartheta) =\frac{1}{\sqrt{2}}\begin{pmatrix}
    e^{in_1^+\vartheta} \varphi_1^+(r) & e^{in_1^-\vartheta} \varphi^-_1(r)&  0 & 0&\cdots \\
    ie^{in_1^+\vartheta} \varphi_1^+(r) &-ie^{in_1^-\vartheta} \varphi^-_1(r)&0 &0 &   \cdots \\
    0&0 & e^{in_2^+\vartheta} \varphi^+_2(r)&e^{in_2^-\vartheta} \varphi^-_2(r)&  \cdots\\
   0 &0 &ie^{in_2^+\vartheta} \varphi^+_2(r)& -ie^{in_2^-\vartheta} \varphi^-_2(r)&  \cdots  \\
    \vdots &\vdots &\vdots & \vdots& \ddots   
\end{pmatrix} \ ,\label{ansfield}}
where $t^a$s are the generators of the Cartan subalgebra of $SO(2N)$. 
The conditions for the fields at $r\to\infty$
are fixed by the 
requirement of finite energy configurations:
 \be{\varphi^\pm_a(\infty)=\sqrt{\frac{\xi}{2N}} \ , }
\be{n_a^\pm=  n^{(0)} \mp    n^{(a)} \ , \qquad n^{(0)} \equiv
  \frac{1}{\sqrt{2}}\, h_0(\infty)\ ; 
\quad  n^{(a)} \equiv   \frac{1}{\sqrt{2}} \, h_a(\infty) \ , \label{c1}}
where $n^{(0)}$  and $ n^{(a)}$  are the winding numbers  with respect to the $U(1)$ and to the $a$-th Cartan $U(1)\in SO(2N)$ defined in Eq.~(\ref{cartan}).

Clearly 
\beq \nzero   \equiv n_a^++n_a^-  = 2\, n^{(0)} \ ,  \eeq
 is independent of $a$.
The regularity of the fields requires that the $Q_{A}$s come back to their original value after a $2 \pi$ rotation, and this yields  the quantization condition, 
\beq n_a^{\pm}\in   {\mathbb {Z}} \ ,    \qquad  \forall a \ ,  \eeq 
 implying that {\it  the $U(1)$ winding numbers  $n^{(0)}$  and $ n^{(a)}$   are quantized in half integer units,}  consistently with  
considerations based on the fundamental groups (see Appendix \ref{homotopy} and below).

We need only the information contained in Eqs.~(\ref{aaa}),(\ref{c1}) to evaluate the tension for a BPS solution:
\be{T = 2\, \eta    \, \xi\, \lim_{r\rightarrow\infty}  \int
  d\vartheta \,r \, A^0_\vartheta(r)=2\,\sqrt{2}\,\pi\,\eta\, \xi\,
  h_0(\infty)= 2 \pi  \, \eta \,  \xi \, \nzero= 2 \pi \,  \xi \, |\nzero|\ .    \label{tension}}
The last equality comes from the requirement for the tension to be positive, so $\eta=\mathrm{sign}(\nzero)$. 
Note that the tension depends only on $|\nzero|$, which is twice
 the $U(1)$ winding.

From the BPS equations we obtain the differential equations for the profile functions $h_0$, $h_a$, $\varphi^\pm_a$: 
\begin{align}
\frac{dh_0}{dr}&=- 2\sqrt{2}\,\eta \,  g_1^2 \, r  \,
\left(\sum_a\left(|\varphi_a^+|^2+|\varphi_a^-|^2\right)-\xi\right) \
, \\
 \frac{dh_a}{dr}&=  2 \sqrt{2} \,\eta\,  g_{2N}^2\, r \,
 \left(|\varphi_a^+|^2-|\varphi_a^-|^2\right) \ , \\
 \frac{d\varphi_a^\pm}{dr}&=\eta\, \left(n_a^\pm-\frac{h_0\mp
   h_a}{\sqrt{2}}\right)\frac{\varphi_a^\pm}{r} \ .
\end{align}
In order to cast them in a simple form, we define
$f_0=h_0-\frac{\nzero}{\sqrt{2}}$ and
$f_a=h_a+\frac{n_a^+-n^-_a}{\sqrt{2}}$ and obtain
\begin{align}
\frac{df_0}{dr}&=-2\sqrt{2}\, \eta\,
 g_1^2r\left(\sum_a\left(|\varphi_a^+|^2+|\varphi_a^-|^2\right)-\xi\right)\label{dh0}\ ,\\
\frac{df_a}{dr}&=2\sqrt{2} \, \eta\,
 g_{2N}^2r\left(|\varphi_a^+|^2-|\varphi_a^-|^2\right) \ , \label{dha} \\
\frac{d\varphi_a^\pm}{dr}&=-\eta\, \left(\frac{f_0\mp
 f_a}{\sqrt{2}}\right)\frac{\varphi_a^\pm}{r} \ . \label{dp}
\end{align}
The boundary conditions at $r\to\infty$ are 
\be{\varphi_a^\pm(\infty)=\sqrt{\frac{\xi}{2N}}\ , \quad
  f_0(\infty)=f_a(\infty)=0 \ ,\label{bi}}
There are also regularity conditions at $r=0$ for the gauge fields
$h_0(0)=h_a(0)=0$ which are
\be{f_0(0)=-\frac{\nzero}{\sqrt{2}} \ ,\quad
  f_a(0)=\frac{n_a^+-n_a^-}{\sqrt{2}} \ , \label{r0}}
Solving Eq.~(\ref{dp}) for small $r$ with the conditions (\ref{r0}), we obtain $\varphi_a^\pm\sim r^{n_a^{\pm} \eta} $. To avoid a singular behavior for these profile functions we need
\be{\mathrm{sign}\left(n_a^\pm\right)=\eta \ . \label{c2}}
This condition is consistent with $\eta=\mathrm{sign}(\nzero)$.
 With this condition there are no singularities at $r=0$ and the equations (\ref{dh0}),(\ref{dha}),(\ref{dp}) can be solved numerically with boundary conditions (\ref{bi}),(\ref{r0}).
 
\begin{figure}
\begin{center}
\includegraphics[width=3.1in]{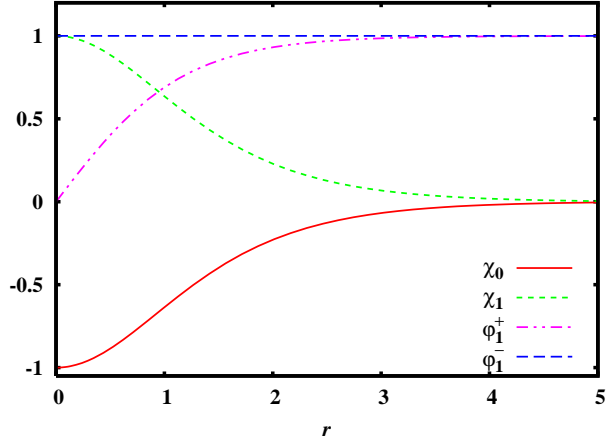}
\caption{\footnotesize Numerically integrated minimum vortex solution
  with $N_0 = 1$, 
   where we have taken the couplings to be $4g_1^2 =
  4g_{2N}^2 = 1$. ($\chi_{i}\equiv \sqrt{2} f_{i}$).}
\label{Vort1}
\end{center}
\end{figure}

\begin{figure}
\begin{center}
\mbox{
  \subfigure{\resizebox{!}{2.3in}{\includegraphics{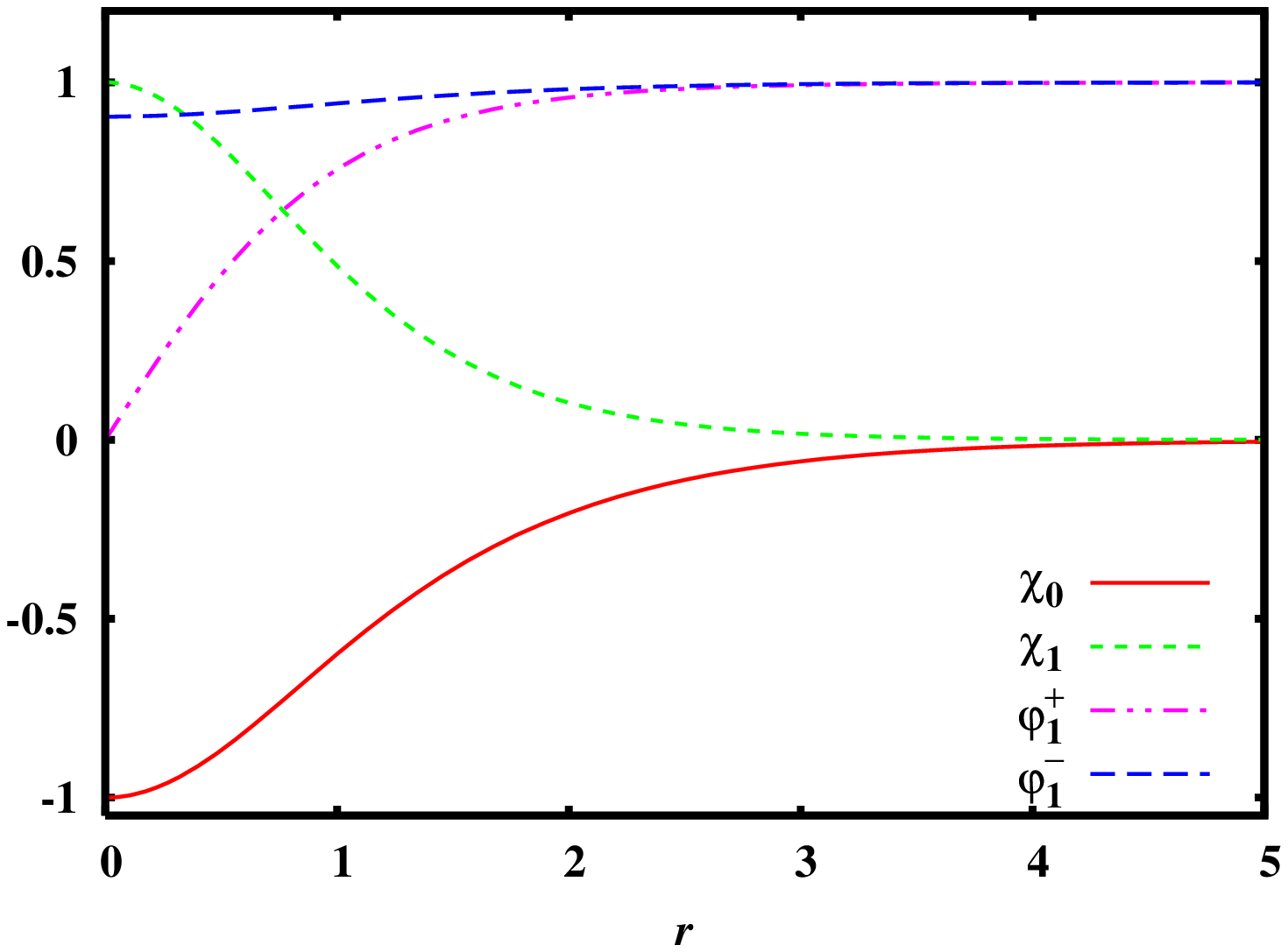}}}\quad
  \subfigure{\resizebox{!}{2.3in}{\includegraphics{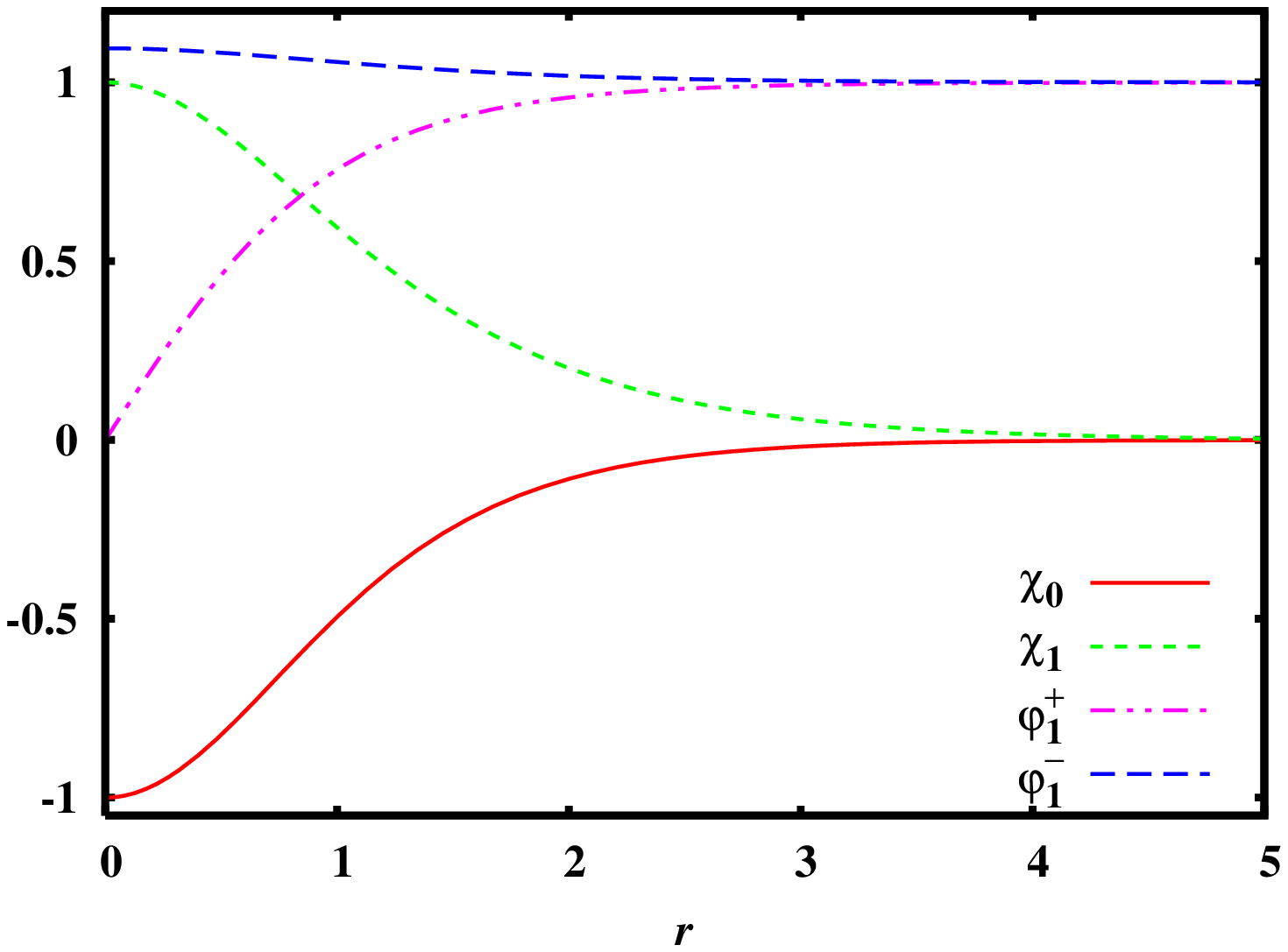}}}}
\caption{\footnotesize Numerically integrated minimum vortex solution
  with $N_0 = 1$, 
 where we have taken the couplings to be $4g_1^2 = 1$ and
  $4g_{2N}^2 = 2$ for the left panel and $4g_1^2 = 2$ and
  $4g_{2N}^2 = 1$ for the right panel. ($\chi_{i}\equiv \sqrt{2} f_{i}$).}
\label{Vort1aneb}
\end{center}
\end{figure}

The profile functions for the simplest vortex $\nzero=1,\ n^+_1=1,\ n^-_1=0$ in the $SO(2)\times U(1)$ theory are shown in Figure~\ref{Vort1}, \ref{Vort1aneb}. The profile functions ($f_{0}, f_{a}, \varphi_{a}^{+},  \varphi_{a}^{-}$) for the minimal 
 vortex $N_{0}=1$, $n_{i}^{+}=1$,  $n_{i}^{-}=0$ in the $SO(2N)\times U(1)$ theory can be obtained by rescaling $g_{2N}^2\rightarrow g_{2N}^2/N$ and then taking all $\varphi_{a}^{\pm}$ equal to  the profile functions shown above rescaled by a factor $1/\sqrt{N}$. Similarly, solutions corresponding to the exchange $(n_{a}^{+},n_{a}^{-})=(1,0)\leftrightarrow(0,1)$ can be obtained by exchanging $f_a\leftrightarrow-f_a$ and $ \varphi_{a}^{+} \leftrightarrow \varphi_{a}^{-}$. The typical length scale of the profile functions is $1/\sqrt{\xi}$, which is the only dimensional parameter in the Bogomol'nyi equations.

 

%

\subsection{Vortex moduli space}

To study the space of solutions of the BPS equations  we have obtained above,  it is convenient to rewrite the ansatz (\ref{ansfield}) for the squark fields in the original basis:
\begin{align}
q(r,\vartheta)&=\begin{pmatrix}
  \mathbf{M}_1(r,\vartheta) & 0 &0& \cdots \\
   0 & \mathbf{M}_2(r,\vartheta) & 0& \cdots \\
   0 & 0 & \mathbf{M}_3(r,\vartheta) & \cdots \\
    \vdots &\vdots &\vdots &\ddots   
\end{pmatrix}\ , \label{ansfield2} \\
\mathbf{M}_a(r,\vartheta)&=\frac{1}{2}\begin{pmatrix}
    e^{in_a^+\vartheta} \varphi_a^+(r) + e^{in_a^-\vartheta} \varphi^-_a(r)& -i\left(e^{in_a^+\vartheta} \varphi_a^+(r) - e^{in_a^-\vartheta} \varphi^-_a(r)\right)\\
     i\left(e^{in_a^+\vartheta} \varphi_a^+(r) - e^{in_a^-\vartheta} \varphi^-_a(r)\right) & e^{in_a^+\vartheta} \varphi_a^+(r) + e^{in_a^-\vartheta} \varphi^-_a(r)
\end{pmatrix} \ . \nonumber 
\end{align}
In this basis the action of the $SO(2N)_{C+F}$ transformations on squark fields is simply $q'= O \, q \, O^T$. The first observation is that if $\hat{q}(r,\vartheta)$ is a solution to the BPS equations, $O \, \hat{q}(r,\vartheta)\,  O^T$ is also a solution. 
Note also that these  solutions  are physically distinct because they
are related by a global symmetry. In this way, from a single solution
of the form (\ref{ansfield}), we can obtain a whole continuous
$SO(2N)$ orbit of solutions.  Any given vortex solution is a point
 in the moduli space and $SO(2N)_{C+F}$ acts as an isometry on this space. 

From Eqs.~(\ref{c1}) and (\ref{c2}), we see that regular solutions are described by a set of $2N+1$ integers $\nzero$,$n_a^\pm$ which satisfy the following conditions:
\begin{eqnarray}
& n_a^++n_a^-=\nzero \ , \quad \forall a \ , \\
& \mathrm{sign}(n_a^+)=\mathrm{sign}(n_a^-)=\mathrm{sign}(\nzero) \ ,
  \quad \forall a \ ,
\end{eqnarray}
where $\nzero\in \mathbb{Z}$ 
 is related to the winding around the $U(1)$ and is the only parameter of the solution which enters the tension $T=2\pi\xi |\nzero|$.

Let us  study the solutions with the minimum tension. Minimal vortices have $\nzero=\pm 1$ and $T=2\pi\xi$. Note that solutions with $\nzero <0$ can be obtained by taking the complex conjugate of solutions with $\nzero >0$, so from now on we will consider only solutions with  positive $\nzero$.   These vortices can be divided into two groups, the first has $2^{N-1}$ representative
(basis)  vortices which are  
\begin{equation}\nzero=1,  \qquad\begin{pmatrix}
  n_1^+ & n_1^-  \\
  n_2^+ & n_2^-  \\
   \vdots & \vdots \\
   n_{N-1}^+ & n_{N-1}^- \\
n_N^+ &n_N^-
\end{pmatrix}
=\begin{pmatrix}
  1 & 0  \\
  1 & 0  \\
   \vdots & \vdots \\
   1 & 0 \\
1 & 0
\end{pmatrix}, \,  
\begin{pmatrix}
  0 & 1  \\
  0  & 1  \\
   1 & 0  \\
   \vdots & \vdots \\
   1 & 0 
\end{pmatrix},  
\, 
\ldots,  \label{minimal1}
\end{equation}
which all have an even number of  $n_{i}^{-}$'s equal to  $1$;  and the second set is represented  by  $2^{N-1}$ vortices,  characterized by
the integers
\begin{equation}\nzero =1,  \qquad\begin{pmatrix}
  n_1^+ & n_1^-  \\
  n_2^+ & n_2^-  \\
   \vdots & \vdots \\
   n_{N-1}^+ & n_{N-1}^- \\
n_N^+ &n_N^-
\end{pmatrix}
=\begin{pmatrix}
  1 & 0  \\
  1 & 0  \\
   \vdots & \vdots \\
   1 & 0 \\
0 & 1
\end{pmatrix},\,
 \begin{pmatrix}
  1 & 0  \\
   \vdots & \vdots \\
    1 & 0  \\
   0 & 1   \\
   1 & 0 
\end{pmatrix},\ldots,
\label{minimal2}
\end{equation}
with an odd number of  $n_{i}^{-}$'s equal to  $1$.

These two sets belong to two distinct  orbits of $SO(2N)_{C+F}$.
To see this one must study the way they transform under  $SO(2N)_{C+F}$.  Consider for instance the case of  $N=2$: the $SO(4)_{C+F}$ transformations
 $\begin{pmatrix}
  \sigma_3 & 0  \\
0 & \sigma_3   
\end{pmatrix}$ and $\begin{pmatrix}
  0 & -\mathbf{1} \\
   \mathbf{1} & 0   
\end{pmatrix}$ exchange $(n_1^+,n_2^+)\leftrightarrow (n_1^-,n_2^-)$ and $(n_1^+,n_1^-)\leftrightarrow (n_2^+,n_2^-)$, respectively.
 In the general $SO(2N)$ case, two solutions differing
 by the exchange $(n_i^+,n_j^+)\leftrightarrow (n_i^-,n_j^-)$ or $(n_i^+,n_i^-)\leftrightarrow (n_j^+,n_j^-)$ for some $i$,$j$, therefore belong to the same orbit of $SO(2N)_{C+F}$. The vortices in the set (\ref{minimal1}) 
 belong to a  continuously degenerate set of minimal vortices;  the set (\ref{minimal2})  form the ``basis'' of another, 
 degenerate set. The two sets do not mix under the $SO(2N)$ transformations. 
 
 In order to see better what these two sets might represent, and to see how each vortex transforms under $SO(2N)_{C+F}$,  let us assign  the two ``states'',  $|{\uparrow}\ket_{j}$,  $ |{\downarrow}\ket_{j}$   of a $j$-th ($\frac{1}{2}$)  spin,   $j=1,2,\ldots, N$,  to  the pair of vortex winding numbers  $(n_{j}^{+}, n_{j}^{-})=(0,1), (1,0)$.   Each of the $2^{N}$   minimum vortices  (Eqs.~(\ref{minimal1}),(\ref{minimal2}))   can then be represented by  the  $2^{N}$  spin state, 
\beq      |  s_{1}  \ket  \otimes  |  s_{2}  \ket  \otimes \cdots |
s_{N}  \ket\ ,  \qquad  |  s_{j}  \ket =  |{\uparrow}\ket=(0,1)\ ,
\quad {\rm or}  \quad |{\downarrow}\ket =(1,0)\ .
\label{Nspin} \eeq
For instance the first vortex of Eq.~(\ref{minimal1})  corresponds to the state,  
$ |  {\downarrow}   {\downarrow}  \ldots  {\downarrow}  \ket$.

Introduce now the ``gamma matrices'' as  direct products of $N$ Pauli matrices acting as
\begin{align}
\gamma_{j} &\equiv \underbrace{\tau_{3}\otimes \cdots \otimes
  \tau_{3} }_{j-1} \otimes \tau_{1} \otimes {\mathbbm {1}} \otimes
\cdots  \otimes  {\mathbbm {1}}\ , \qquad    (j=1,2,\ldots, N)\ ; \\
\gamma_{N+j} &\equiv \underbrace{\tau_{3}\otimes \cdots \otimes
  \tau_{3} }_{j-1} \otimes \tau_{2} \otimes {\mathbbm {1}} \otimes
\cdots  \otimes  {\mathbbm {1}}\ , \qquad   (j=1,2,\ldots, N)\ . 
\end{align}
$\gamma_{k}$, $k=1,2,\ldots, 2N$  satisfy the  Clifford algebra
\[  \{\gamma_{i}, \gamma_{j} \} = 2 \, \eta_{ij}\ , \qquad
i,j=1,2,\ldots, 2N \ ,   \]
and the   
$SO(2N)$ generators can accordingly be constructed  by $\Sigma_{ij}=   \frac{1}{4 i }  [\gamma_{i}, \gamma_{j}]$. 
$SO(2N)$ transformations (including finite transformations)  among the vortex solutions can thus be represented
by the transformations among the  $N$-spin states, (\ref{Nspin}).

As each of  $\Sigma_{ij}$  $(i \ne j)$  flips exactly two spins, the two sets  (\ref{minimal1})  and  (\ref{minimal2}) clearly belong to
two distinct orbits of $SO(2N)$.   In fact, 
 a ``chirality''   operator   
\beq  \Gamma_{5} \equiv   P \, \prod_{j=1}^{2N} \,  \gamma_{j}\
     ,\qquad   \{ \Gamma_{5},  \gamma_{j}\}=0\ , \quad j=1,2,\ldots,
     2N \ ,   
\eeq
anticommutes with all $\gamma_{j}$'s,  where $P=1$ ($N$ even) or  $P=i$  ($N$ odd), hence commutes with $SO(2N)$.   
The two sets Eq.~(\ref{minimal1}),  Eq.~(\ref{minimal2})  of minimal vortices thus are seen to transform as two spinor representations 
of  definite chirality, $1$ and $-1$,  respectively (with multiplicity $2^{N-1}$ each).  
%

Every minimal solution is invariant under a $U(N)$ group embedded in $SO(2N)_{C+F}$. This can be seen from the form of the first solution in (\ref{minimal1}) in the basis (\ref{ansfield2}):
\be{q_{(1)}=
f_+(r,\vartheta)\begin{pmatrix} \mathbf{1} && \\ &\ddots & \\ & & \mathbf{1} \end{pmatrix}+
f_-(r,\vartheta)\begin{pmatrix} \sigma_2 && \\ &\ddots & \\ & & \sigma_2 \end{pmatrix}. \label{form}}
This solution is invariant under the subgroup $U(N) \subset SO(2N)$ acting as $U \,q_{(1)}\, U^{T}$, where $U\in U(N)$ commutes with the second matrix in (\ref{form}).

In the $N$-spin state representation above, the vortex (\ref{form})
corresponds to the state with all spins down,   $ |  {\downarrow}   {\downarrow}  \ldots  {\downarrow}  \ket$.    In order to see how the $N$-spin states transform under $SU(N)\subset SO(2N)$, construct the creation and annihilation operators 
\[   a_{j}= \frac{1}{2} (\gamma_{j} - i\, \gamma_{N+j} )\ ; \qquad
a_{j}^{\dagger}= \frac{1}{2} (\gamma_{j} +  i\, \gamma_{N+j})\ ,
\]
satisfying the algebra,
\[   \{a_{j}, a_{k}\}= \{a_{j}^{\dagger}, a_{k}^{\dagger}\}=0\ ,
\qquad \{a_{j}, a_{k}^{\dagger}\}=\delta_{jk}\ .
\]
$SU(N)$ generators acting on the spinor representation, can be constructed as \cite{Georgi} 
\[  T^{a} = \sum_{j,k} \,  a_{j}^{\dagger}  \, (t^{a})_{j k} \, a_{k}\
, 
\]
where $t^{a}$ are the standard $N\times N$ $SU(N)$ generators in the
fundamental representation.    The state $ |  {\downarrow}
{\downarrow}  \ldots  {\downarrow}  \ket$  is clearly annihilated by all  $T^{a}$, as  it is annihilated by all 
\[   a_{k} =\underbrace{\tau_{3}\otimes \cdots \otimes
  \tau_{3} }_{k-1} \otimes \tau_{-}  \otimes {\mathbbm {1}} \otimes
\cdots  \otimes  {\mathbbm {1}}\ , \qquad k=1,2,\ldots N:
\]
thus, the vortex (\ref{form}) leaves $U(N)$ invariant.

All other solutions can be obtained as $R\, q_{(1)}\, R^T$ with $R \in O(2N)$, so each solution is invariant under an appropriate $U(N)$ subgroup $R\, U\, R^T$. 
This means that the moduli space contains two copies of the coset space 
\beq {\cal M}=  {SO(2N) / U(N)} \ .
\eeq  
The points in each coset space transform according to a spinor representation of definite chirality, each with dimension $2^{N-1}$. 
 When discussing the topological properties of vortices, we will see that these disconnected parts correspond to different elements of the homotopy group.

Vortices of higher windings are described by $\nzero>1$. 
In the simplest non-minimal case, the vortices are described by:
\begin{equation}\nzero=2\ , \qquad
\begin{pmatrix}
  2 & 0  \\
  2 & 0  \\
   \vdots & \vdots \\
   2 & 0 \\
2 & 0
\end{pmatrix} , \begin{pmatrix}
  2 & 0  \\
  2 & 0  \\
   \vdots & \vdots \\
   2 & 0   \\
   0 & 2 
\end{pmatrix}, \begin{pmatrix}
  2 & 0  \\
  2 & 0  \\
   \vdots & \vdots \\
   2 & 0   \\
   1 & 1 
\end{pmatrix} \ldots \begin{pmatrix}
  2 & 0  \\
  1 & 1  \\
   \vdots & \vdots \\
   1 & 1   \\
   1 & 1 
\end{pmatrix}, \begin{pmatrix}
   1 & 1  \\1 & 1  \\
   \vdots & \vdots \\
   1 & 1   \\
   1 & 1 
\end{pmatrix}. \label{m2}
\end{equation}
These orbits correspond to parts of the moduli space whose structure corresponds to the coset spaces 
$SO(2N)\diagup U(N-k)\times SO(2k)$,
where $k$ is the number of $(1,1)$ pairs.   Analogously vortices with $\nzero \ge 3 $ can be constructed.

The argument that the minimum vortices transform as two spinor
representations  implies that the $\nzero=2$ vortices (\ref{m2}) transform as 
various irreducible antisymmetric tensor representations of $SO(2N)_{C+F}$, appearing in the decomposition of  products of two     spinor representations: e.g. 
\beq   2^{N-1} \otimes   2^{N-1} \ \mathrm{or}\ 2^{N-1} \otimes   \overline{2^{N-1}} \ ,  \label{prodspinors} \eeq 
Although all these vortices are degenerate in the semi-classical  approximation,  non-BPS  corrections will lift the degeneracy, leaving only the degeneracy among the vortices transforming as an irreducible multiplet of the group $SO(2N)_{C+F}$.  For instance  the last vortex  $n_{a}^{+}= n_{a}^{-}=1$, for all $a$, carries only the unit $U(1)$ winding and is a 
singlet, the second last vortex and analogous ones    
belong to a $\underline{\mathbf{2N}}$, and so on.  

 Due to the fact that the tension depends only on $\nzero= 2\, n^{(0)}$   (twice the $U(1)$  winding)  the degeneracy pattern of the vortices 
does not simply reflect the homotopy map  which relates the vortices to the massive monopoles. 
The monopole-vortex correspondence will be discussed  in Section~\ref{corres} below. 

\begin{figure}
\begin{center}
\mbox{
  \subfigure{\resizebox{!}{2.3in}{\includegraphics{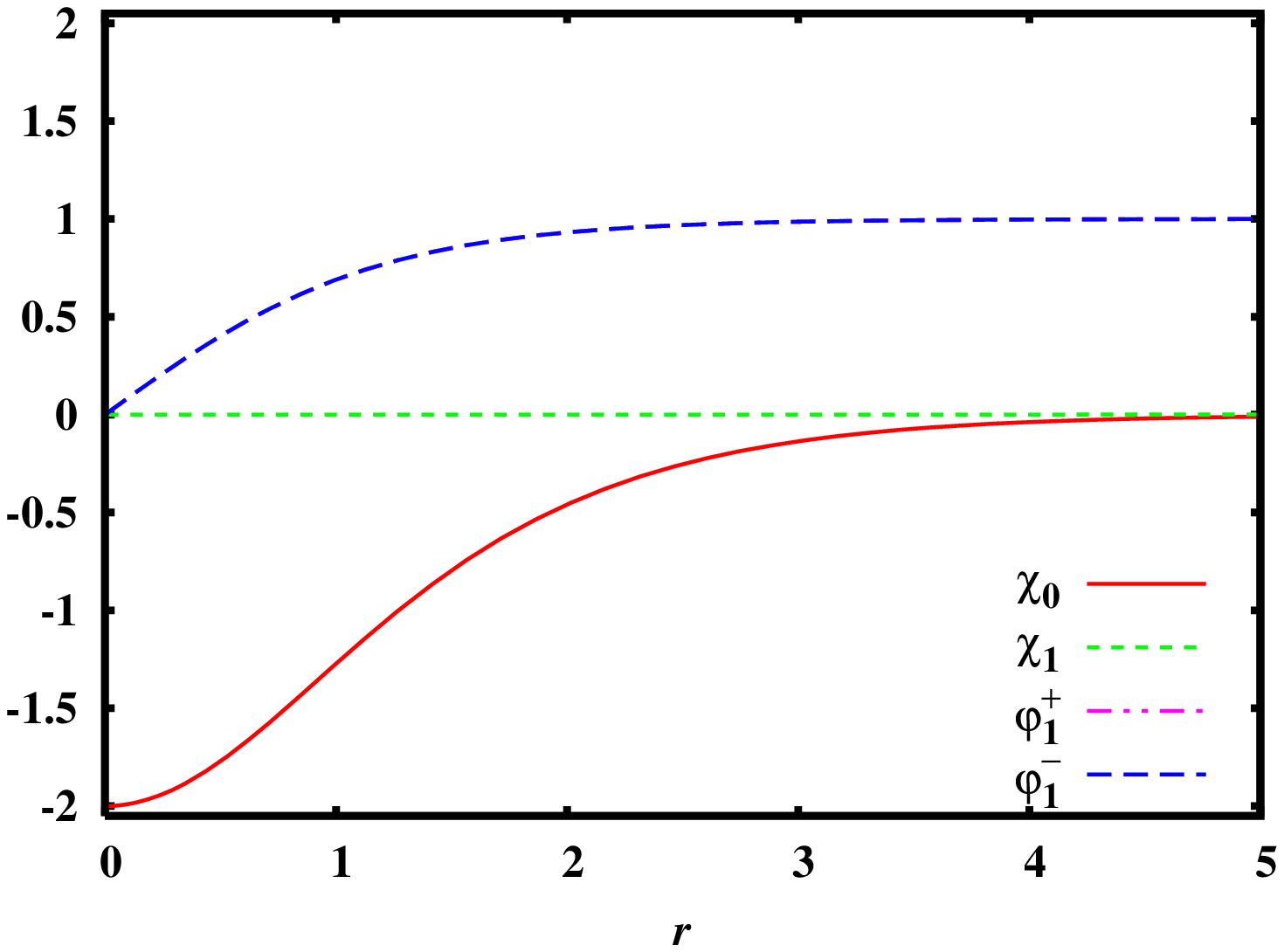}}}\quad
  \subfigure{\resizebox{!}{2.3in}{\includegraphics{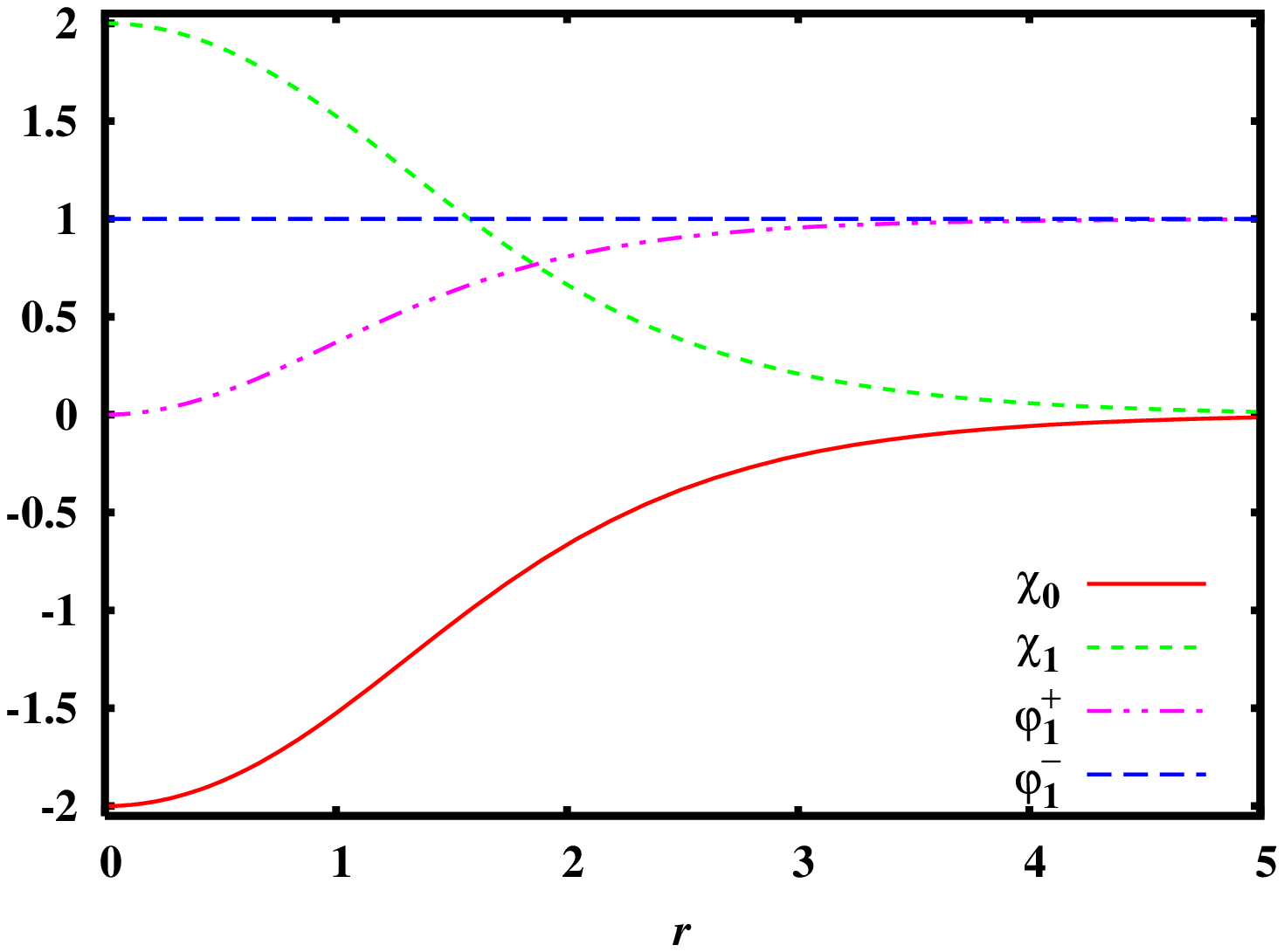}}}}
\caption{\footnotesize Numerically integrated minimum vortex solution
  with $N_0 = 2$, where we have taken the couplings to be $4g_1^2 =
  4g_{2N}^2 = 1$. In the left panel we have shown the element $(n^+,
  n^-) = (1,1)$ and in the right panel the element  $(n^+,
  n^-) = (2,0)$. The dependence of the couplings turns out to be similar
   to the case of the minimal vortex $(n^+,n^-) =
  (1,0)$. ($\chi_{i}\equiv \sqrt{2} f_{i}$).}
\label{Vort1nm}
\end{center}
\end{figure}

The profile functions $(f_0,f_a,\varphi_a^+,\varphi_a^-)$ for the
simplest non-minimal vortex, $N_0 = 2$ are illustrated in Figure
\ref{Vort1nm}. 
In the figure is just considered the two simplest elements $(n^+,n^-)
= (1,1)$ and $(n^+,n^-) = (2,0)$. Adding elements of the same type
corresponds just to a rescaling of the coupling $g_{2N}^2$ and of the functions $\varphi_a^{\pm}$ as in the
minimal vortex case ($N_0 = 1$). Adding elements of different types ($(2,0)$ or $(1,1)$)
does not induce new behavior. 

\section{Vortices in $SO(2N+1)$ theories}\label{Modd}

Consider now the case of a theory with symmetry breaking
\begin{equation} SO(2N+3) \stackrel{v_1}{\longrightarrow} SO(2N+1)
  \times U(1) \stackrel{v_2}{\longrightarrow} {\mathbbm {1}}\ .    \label{hierarchyodd} \end{equation} 
The fields which remain massless after the first symmetry breaking can be found exactly as in the even $SO$ theories
by use of various $SO(3)$ groups, leading to Eq.~(\ref{Expanded}),  with  $A=1,2,\ldots, N_{f}$ where we now take $N_{f}= 2 N+1$. 
The light quarks can get color-flavor locked VEVs  as in Eq.~(\ref{squarkVev}), leading to a vacuum with global  $SO(2N+1)_{C+F}$
symmetry.   

The ansatz (\ref{ansfield2}) must be modified as follows
\be{q(r,\vartheta)=\begin{pmatrix}
  \mathbf{M}_1(r,\vartheta) & \cdots &0& 0 \\
   \vdots & \ddots & \vdots& \vdots\\
   \vdots & \cdots & \mathbf{M}_N(r,\vartheta) & 0 \\
    0 & \cdots &0 & e^{i\hat{n}\vartheta}\hat{\varphi}(r)   
\end{pmatrix}\ ,\label{ansfield3}}
introducing a new integer  $\hat{n}$ and a new profile function $\hat{\varphi}(r)$.
The equation (\ref{dh0}) becomes 
\be{\frac{df_0}{dr}=-   2 \, \sqrt{2}\, \eta  \, 
  g_1^2r\left(\sum_a\left(|\varphi_a^+|^2+|\varphi_a^-|^2\right)+|\hat{\varphi}|^2-\xi\right)\ ,}
while the condition of finite energy gives
\begin{align}
\hat{\varphi}(\infty)&=\sqrt{\frac{\xi}{2N+1}}\ ,\\
\hat{n}&=\frac{h_0(\infty)}{\sqrt{2}}=\frac{N_0}{2}\ ,\label{hatnint}
\end{align}
and the equation for $\hat{\varphi}(r)$ is
\be{\frac{d\hat{\varphi}}{dr}=\eta  \, \left(\hat{n}-\frac{h_0}{\sqrt{2}}\right)\frac{\hat{\varphi}}{r}=-\eta  \, \frac{f_0}{\sqrt{2}}\frac{\hat{\varphi}}{r}\
.}

Note that the condition (\ref{hatnint}) fixes $\hat{n}$ in terms of $N_0$:   as $\hat{n}$  must be an integer,  this theory contains only vortices with even $N_0$. This can be traced to the different structure of the gauge groups. In fact, $SO(2N+3)$ has no center, so the pattern of symmetry breaking is 
\be{SO(2N+3)\rightarrow SO(2N+1)\times U(1)\rightarrow {\mathbbm {1}}\
  ,  }
and there are no vortices with half-integer winding around the $U(1)$, or around any other Cartan $U(1)$ subgroups.

The vortices are classified by the same integers $n_a^\pm$ as before, but now there are $SO(2N+1)_{C+F}$ transformations which exchange $n_a^+ \leftrightarrow n_a^-$ singly.  The minimal vortices are labeled   by 
\begin{equation}    (n_{a}^{+}, n_{a}^{-})=
\begin{pmatrix}
  2 & 0  \\
  2 & 0  \\
   \vdots & \vdots \\
   2 & 0 \\
2 & 0
\end{pmatrix},  \begin{pmatrix}
  2 & 0  \\
  2 & 0  \\
   \vdots & \vdots \\
   2 & 0   \\
   1 & 1 
\end{pmatrix} \ldots \begin{pmatrix}
  2 & 0  \\
  1 & 1  \\
   \vdots & \vdots \\
   1 & 1   \\
   1 & 1 
\end{pmatrix}, \begin{pmatrix}
   1 & 1  \\1 & 1  \\
   \vdots & \vdots \\
   1 & 1   \\
   1 & 1 
\end{pmatrix},   \qquad  {\hat n}=1\ .\label{vorodd} 
\end{equation}
The moduli space contains subspaces corresponding to these orbits, whose structure is that of the coset spaces 
$SO(2N+1)\diagup \left(U(N-k)\times SO(2k+1)\right)$
where $k$ is the number of $(1,1)$ pairs.

\begin{figure}
\begin{center}
\mbox{
  \subfigure{\resizebox{!}{2.3in}{\includegraphics{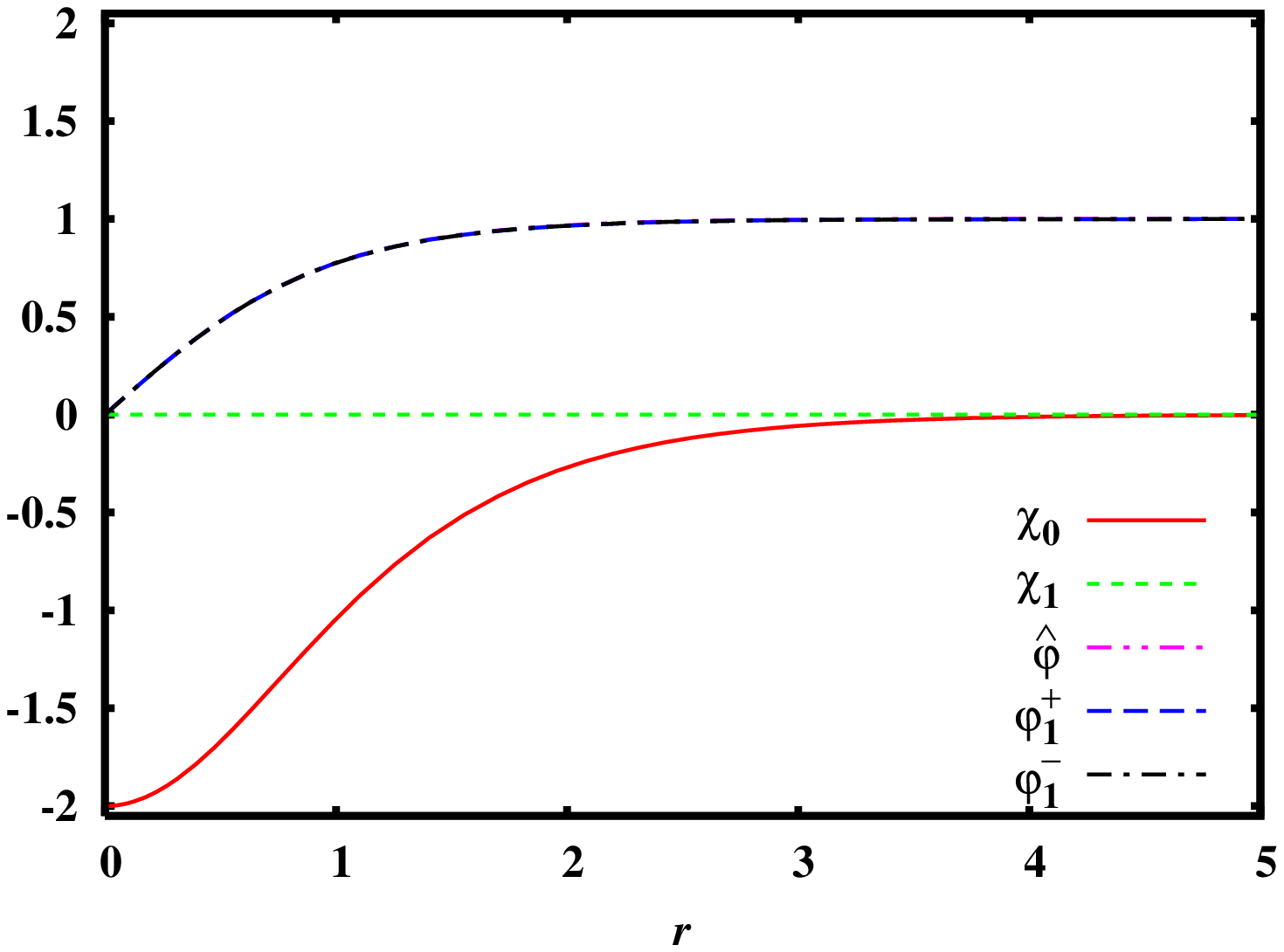}}}\quad
  \subfigure{\resizebox{!}{2.3in}{\includegraphics{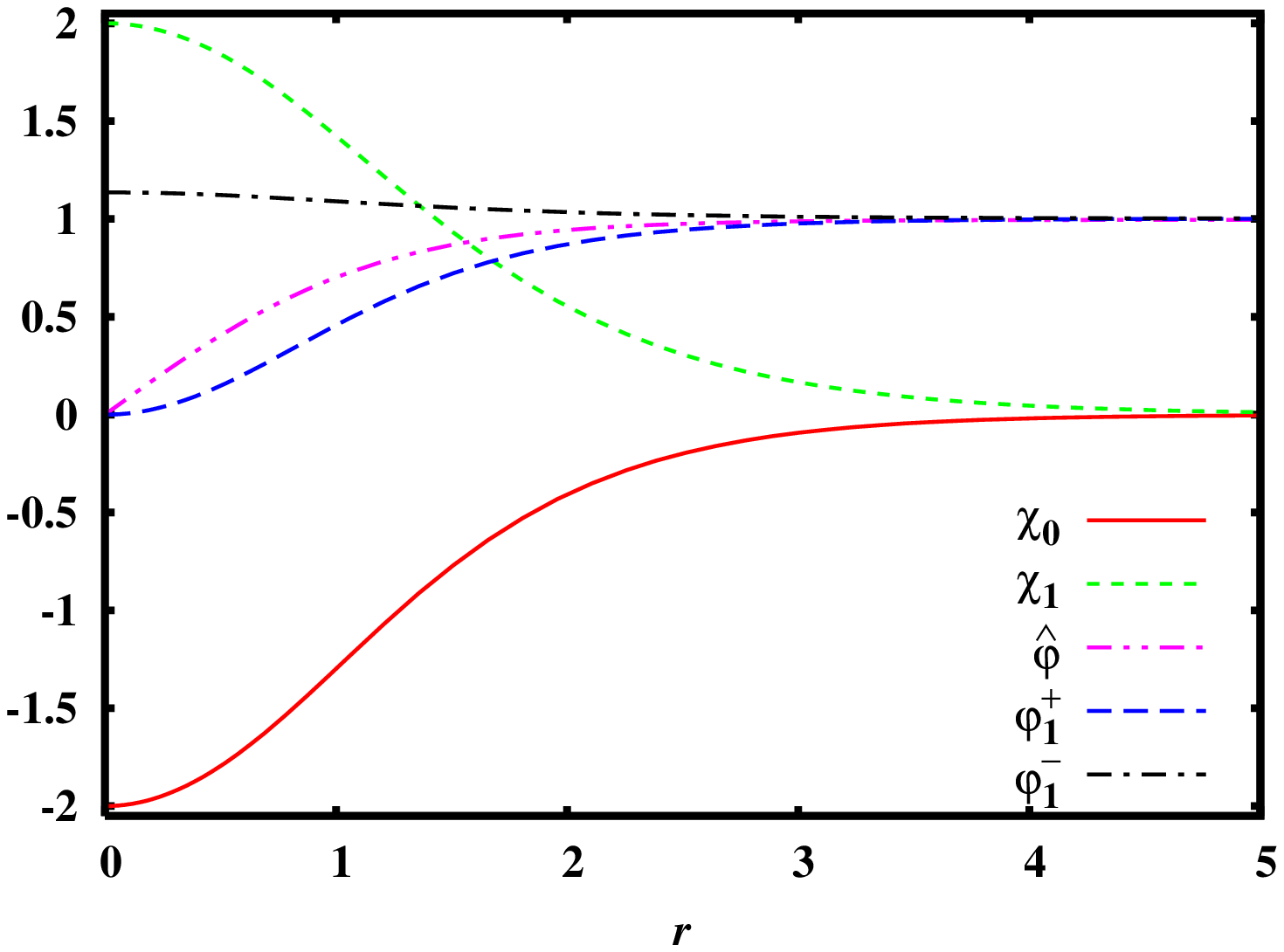}}}}
\caption{\footnotesize Numerically integrated minimum vortex solution of the
  $SO(2N+1)$ theory, with $N_0 = 2$ and we take the couplings to be
  $4g_1^2 = 4g_{2N}^2 = 1$. In the left panel we have $(n_{1}^{+},
  n_{1}^{-})=(1,1)$ and in the right panel $(n_{1}^{+},
  n_{1}^{-})=(2,0)$. The dependence of the couplings turns out to be
   analogous to the case of the $(n_{1}^{+},
  n_{1}^{-})=(1,0)$  vortex. ($\chi_{i}\equiv \sqrt{2} f_{i}$).}
\label{Vort2}
\end{center}
\end{figure}

The vortex profile functions are shown in Figure \ref{Vort2}. 

\section{Monopoles, vortices,  topology and confinement\label{corres}}

\subsection {Homotopy map \label{sec:mapping}} 

The multiplicity of vortex solutions depends on the particular topology of the symmetry-breaking pattern of our model. 

Usually, in systems with a gauge Lie group $G$ and a symmetry-breaking pattern
\begin{equation} G \stackrel{v_1}{\longrightarrow} H
  \stackrel{v_2}{\longrightarrow} {\mathbbm {1}} \ ,   \end{equation} 
there are:
\begin{itemize}
\item Stable Dirac monopoles, classified by $\pi_1(G)$; 
\item Regular monopoles, classified by $\pi_2(G/H)$; topologically stable only in the limit $v_2 \rightarrow 0$; 
\item Vortices, classified by $\pi_1(H)$; if they correspond to a non-trivial element of $\pi_1(G)$, they are topologically stable; otherwise they are topologically stable only in the limit $v_1 \rightarrow \infty$. 
\end{itemize}
Monopoles and vortices are related by the topological correspondence \cite{ABEKM}
\be{\pi_2(G/H)=\pi_1(H)/\pi_1(G) \ , \label{topol}}
so regular monopoles correspond to vortices which are trivial with respect to $\pi_1(G)$, while vortices which are non-trivial with respect to $\pi_1(G)$ correspond to Dirac monopoles. 

In our theories of type $D_{N}$, however, the center $C_G  = {\mathbb {Z}}_{2}$ acts trivially on all fields and the breaking pattern is
\be{G \stackrel{v_1}{\longrightarrow}
 H \stackrel{v_2}{\longrightarrow} C_G  \ , \label{instead}}
and the topological relation (\ref{topol}) is not directly  useful. In fact, vortices are classified by $\pi_1(H/C_G)$, which is a richer homotopy group than $\pi_1(H) \sim \pi_2(G/H)\times \pi_1(G)$. In our example the relevant group is
\be{\pi_1\left(\frac{SO(2N)\times U(1)}{{\mathbb
      {Z}}_2}\right)=\mathbb{Z}\times {\mathbb {Z}}_2 \ .}
The failure of (\ref{topol}) would  mean that the correspondence between monopoles and vortices is lost.

Actually,  it is better to formulate the problem as follows. The theory contains only fields in the adjoint representation, so we can  neglect the center $C_G$ from the beginning and consider the gauge group as  $G'=G/C_G$.  In our example, the gauge group of the high-energy theory can be taken as $G'=SO(2N+2)/{\mathbb {Z}}_2$, broken to $H'=(SO(2N)\times U(1))/{\mathbb {Z}}_2$ at scale $v_1$ and then completely broken at scale $v_2$:
\be{G' \stackrel{v_1}{\longrightarrow}
 H' \stackrel{v_2}{\longrightarrow} {\mathbbm{1}} \ .   \label{insteadof}}
instead of  Eq.~(\ref{instead}). 
 Then the relation  (\ref{topol}) reads  
\be{\pi_2\left(\frac{SO(2N+2)}{SO(2N)\times
    U(1)}\right)=\frac{\pi_1\left(\frac{SO(2N)\times U(1)}{{\mathbb
	{Z}}_2}\right)}{\pi_1\left(\frac{SO(2N+2)}{{\mathbb
	{Z}}_2}\right)} \ .}
Regular monopoles are classified by the same homotopy group as before,
because \be{\frac{SO(2N+2)/{\mathbb {Z}}_2}{(SO(2N)\times
    U(1))/{\mathbb {Z}}_2}=\frac{SO(2N+2)}{SO(2N)\times U(1)} \ ,}
while for Dirac monopoles the situation is different: the relevant homotopy group is not {$\pi_1\left(SO(2N+2)\right)$}, but the larger group $\pi_1\left(SO(2N+2)/{\mathbb {Z}}_2\right)$   (see Appendix \ref{homotopy})
\be{\pi_1\left(\frac{SO(4J)}{{\mathbb {Z}}_2}\right)={\mathbb
    {Z}}_2\times {\mathbb {Z}}_2 \ ,  \label{rich1}}
while
\be{\pi_1\left(\frac{SO(4J+2)}{{\mathbb {Z}}_2}\right)={\mathbb {Z}}_4
  \ ,   \label{rich2}}
so that the Dirac monopoles have quantized ${\mathbb {Z}}_2\times {\mathbb {Z}}_2  $  or ${\mathbb {Z}}_4$   charges.

   This means that the theory has a larger set of monopoles, and the correspondence between monopoles and vortices (which confine them) is rather subtle\footnote{Note that the Lagrangian and fields for the two theories with gauge group $SO(2N+2)$ and $SO(2N+2)/{\mathbb {Z}}_2$ are the same. 
The set of vortices is the same for both theories and has a topological correspondence with the larger set of monopoles.}.
   
    In appendix \ref{homotopy} we briefly review the structure of the homotopy groups which are relevant for this analysis.


Finally, for the groups of type  $B_{N}$,  the situation is slightly simpler as there is no non-trivial center.  The non-trival element of  $\pi_{1}\left(SO(2N+3)\right)= {\mathbb {Z}}_{2}$  represents the (unique type of)  Dirac monopoles;   the elements of  $\pi_{1}\left(SO(2N+1)\times U(1)\right)= {\mathbb {Z}}_{2} \times {\mathbb {Z}}\,$ 
label the vortices of the low-energy theory.  The vortices whose (non-trivial) winding in the group  $SO(2N+1)\times U(1)$  corresponds to a contractible  loop in the parent theory, confine the regular monopoles.

\subsection{Flux matching \label{matching}}

To establish the  matching between regular GNO monopoles and low-energy  vortices, we use the topological correspondence discussed in the previous section. Dirac monopoles are  classified by  $\pi_1\left(SO(2N+2)/{\mathbb {Z}}_2\right)$ or  by  $\pi_1\left(SO(2N+3)\right)$  depending on the gauge group, but regular monopoles are classified by $\pi_2\left(\frac{SO(2N+2)}{SO(2N)\times U(1)}\right)$ or by  $\pi_2\left(\frac{SO(2N+3)}{SO(2N+1)\times U(1)}\right)$ , i.e. homotopically non-trivial paths in the low-energy gauge group,  which are trivial in the high-energy gauge group. Regular monopoles can be sources for the vortices corresponding to these paths.

The vortices of the  lowest tension which satisfy this requirement are those with $N_0=\pm 2$ and $\sum_a(n_a^+-n_a^-)/2$ odd, so vortices corresponding to minimal GNO monopoles belong to the $SO(2N)_{C+F}$ orbits classified by (\ref{m2}) with an odd number of $(\pm 2,0)$ pairs.

For a better understanding of this correspondence, we can also use flux matching between vortices and monopoles \cite{ABEK}. There are $2N$ GNO monopoles obtained by different embeddings of broken $SU(2) \subset SO(4)$ in $SO(2N+2)$. In a gauge where $\phi$ is constant, their fluxes are
\be{\int_{\mathcal{S}^2} d\vec{S} \cdot \vec{B}^a t^a=2\sqrt{2}\pi(t_0
  \pm t_i) \ , } 
where $t_0 \pm t_i$ is the unbroken generator of the broken $SU(2)$ subgroup.
In the same gauge, the flux of a vortex is
\be{\int_{\mathbb{R}^2} d^2x B_z^a t^a=-\nzero\sqrt{2}\pi t_0
  +\left(n_j^+-n_j^-\right)\sqrt{2}\pi t_j \ ,}
so the fluxes agree for $\nzero=-2$, $n_j^+-n_j^-=\pm 2 \delta_{ij}$. The antimonopoles correspond to the opposite sign $\nzero=2$.


\subsection {Monopole confinement: the $SO(2N)$ theory}

We have now all the tools needed  to analyze the duality in the $SO$ theories at hand.  The general scheme for mapping the monopoles and vortices has been set up in Section \ref{sec:mapping}.    An important point to keep in mind is that, while 
 the vortex tension depends only on the $U(1)$  flux  in our particular model
(Eq.~(\ref{tension})), the classification of vortices according to the first homotopy group reflects the other Cartan charges (windings in $SO(2N)$ or $SO(2N+1)$). It is necessary to keep track of these 
to see how the vortices in the low-energy theory are associated with the monopoles of the high-energy system.

  First consider the theories of type $D_{N}$, with the symmetry breaking 
\begin{equation} SO(2N+2) \stackrel{v_1}{\longrightarrow} SO(2N)
  \times U(1) \stackrel{v_2}{\longrightarrow} {\mathbbm {1}}\ .    \label{hierarchyBisbis} \end{equation} 
studied in detail in the preceding sections.  The vortices with minimum winding, $N_{0}=1$, of  Eqs.~(\ref{minimal1}), (\ref{minimal2}),  correspond to 
the minimum non-trivial element of  $\pi_{1}\left((SO(2N)\times U(1))/{\mathbb {Z}}_2\right)$, which represent also the minimal  elements of  $\pi_{1}\left(SO(2N+2)/{\mathbb {Z}}_2\right)$.  This last fact means that  they are stable in the full theory. 
They would  confine Dirac monopoles of the minimum charge in the underlying theory, 
$1$ of ${\mathbb {Z}}_{4}$ or  $(1,0)$ or  $(0,1)$  of ${\mathbb {Z}}_{2} \times {\mathbb {Z}}_{2} $, see  Appendix \ref{homso2np2}.   

Consider now the  vortices Eq.~(\ref{m2})  with  $N_{0}=2$.    As the fundamental group of the underlying theory is given by either 
Eq.~(\ref{rich1}) or  Eq.~(\ref{rich2}),  some of the vortices will correspond to non-contractible loops in the  underlying gauge group: they would be related to the Dirac monopoles and not to the regular monopoles. Indeed, consider the last of Eq.~(\ref{m2}):
\beq  \left(\begin{array}{c}n_{a}^{-} \\n_{a}^{+}\end{array}\right)=
\left(\begin{array}{ccccc}  1 & 1 & 1 & \ldots & 1 \\  1  & 1 & 1 &
  \ldots  & 1\end{array}\right)\ .  \label{last}   \eeq
  It is characterized by
the windings $n^{(0)}=1$, $n^{(a)}=0$ for all
$a$.  Thus it is an ANO  vortex of the $U(1)$  theory, with no flux in the $SO(2N)$ part. It corresponds to a $2\pi$ rotation in $(12)$ plane in the original $SO(2N+2)$ group  -- the path $P$ in Appendix~\ref{homoprimo}: it is to be associated with a Dirac monopole of charge $2$.

The vortices of the type
\beq    \left(\begin{array}{ccccc}  0 & 1 & 1 & \ldots & 1 \\  2  & 1
  & 1 & \ldots  & 1\end{array}\right) \ , \label{toget}   \eeq
and analogous ones (with $(2,0)$ or $(0,2)$ appearing in different positions)  are characterized by the two $U(1)$ windings only:   a flux   $n^{(0)}=1$ and one of the Cartan flux of $SO(2N)$, e.g., $n^{(1)}=1$  ($n^{(a)}=0$, $a \ne 1$).  
They correspond to a  simultaneous  $2\pi$ rotations in $(12)$ and in $(34)$ planes
in the gauge group and it represents a contractible loop in the high-energy gauge group. They confine regular monopoles, as can be seen also by the flux matching argument discussed in section~\ref{matching}.

Part of the continuous moduli of these vortex solutions include  
\beq   SO(2N)   \diagup U(1)\times SO(2N-2) \ , 
\label{dimension}\eeq  
 as the individual soliton  breaks $SO(2N)_{C+F}$ symmetry of the system. This space corresponds to the complex quadric surface $Q^{2N-2}(C)$.   
  As these vortices are 
not elementary but composite of the minimal vortices,  determining their correct moduli space structure is not a simple task. 
 
Nevertheless, there are some indications that these correspond to a  vector representation $\underline{\mathbf{2N}}$  of $SO(2N)_{C+F}$, appearing in the decomposition of the product of two spinor representations, Eq.~(\ref{prodspinors}).  In fact,   the vortex Eq.~(\ref{toget})  arises as a product
\beq    \left(\begin{array}{ccccc}  0 & 0 & 0 & \ldots & 0 \\  1  & 1 & 1 & \ldots  & 1\end{array}\right)
\otimes  \left(\begin{array}{ccccc}  0 & 1 & 1 & \ldots & 1 \\  1  & 0 & 0 & \ldots  & 0\end{array}\right):  
 \label{products}   \eeq
i.e.,  a product of  two spinors of the same chirality  if $N$ is odd;  {\it vice versa},  of spinors of opposite chirality if  $N$ is even. This corresponds precisely to the known decomposition rules in $SO(4m+2)$ and $SO(4m)$  groups  (see e.g., \cite{Georgi}, Eq.~(23.40)).

In order to establish that these vortices indeed transform under the $SO(2N)_{C+F}$  as a  ${\underline{\mathbf{2N}}}$  one needs to construct the moduli matrix \cite{Isozumi:2004vg} for these, and  study explicitly how the points in the moduli space transform.  This problem will be studied elsewhere.

It is interesting to note that there seems to be a relation between the transformation properties of monopoles under the dual GNO group ${\widetilde {SO}}(2N)$ and the transformation properties of the corresponding vortices under the $SO(2N)_{C+F}$ group. In fact, vortices transforming as a vector of $SO(2N)_{C+F}$ have precisely the net magnetic flux of  regular monopoles  in $\underline{\mathbf{2N}}$  of   ${\widetilde {SO}}(2N)$, as classified by the GNO criterion. 

Other vortices in Eq.~(\ref{m2})  correspond to various Dirac (singular) or regular monopoles  in different representations 
of $SO(2N)_{C+F}$.  

\subsection {Monopole confinement: the $SO(2N+1)$ theory} 

In the $B_{N}$ theories with the symmetry breaking 
\begin{equation} SO(2N+3) \stackrel{v_1}{\longrightarrow} SO(2N+1)
  \times U(1) \stackrel{v_2}{\longrightarrow} {\mathbbm {1}}\ . \label{hierarchyodd2} \end{equation} 
the minimal vortices of the low-energy theory have $\nzero=2$.  Reflecting the difference of $\pi_{1}$ group of the underlying theory
as compared to the $D_{N}$ cases (${\mathbb {Z}}_{2}$ as compared to  ${\mathbb {Z}}_{2}\times {\mathbb {Z}}_{2}$  or ${\mathbb {Z}}_{4}$), the $N_{0}=1$ vortices (with  half winding  in $U(1)$ and $SO(2N)$)  are absent here. 

The minimal vortices (\ref{vorodd})  again correspond to different homotopic types and to various $SO(2N+1)$ representations. 
The vortex
\beq       \left(\begin{array}{ccccc}  1 & 1 & 1 & \ldots & 1 \\  1  &
  1 & 1 & \ldots  & 1\end{array}\right), \qquad {\hat n}=1\ ,  \label{lastBis}   \eeq
has the $U(1)$  charge  $n^{(0)}=1$ and no charge with respect to $SO(2N+1)$.  It is associated to the non-trivial element of 
$\pi_{1}\left(SO(2N+3)\right)={\mathbb {Z}}_{2}$:  it is stable in the full theory.  Its flux would match that of a Dirac monopole.   This is a singlet of 
$SO(2N+1)_{C+F}$  (its moduli space consists of a point).  

Consider instead the vortices 
\beq    \left(\begin{array}{ccccc}  0 & 1 & 1 & \ldots & 1 \\  2  & 1
  & 1 & \ldots  & 1\end{array}\right), \qquad {\hat n}=1\ ,  \label{regBN}   \eeq
and analogous ones,  having the winding numbers $n^{(0)}=1$,  $n^{(a)}=\pm 1$,  $n^{(b)}=0$,  $b\ne a$, and ${\hat n}=1$.  
 These would correspond to regular monopoles which,  according to GNO classification, are supposed to belong to a ${\underline {\mathbf{2N}}}$ representation of the dual group  $USp(2N)$.   Again, though it is not a trivial task to establish that these vortices do transform as  ${\underline {\mathbf{2N}}}$ of 
such a group,  there are some hints they indeed do so. It is crucial
that the symmetry group (broken by individual soliton
vortices) is   $SO(2N+1)$:   it is  in fact 
possible to identify the ${\underline {\mathbf{2N}}}$ generators constructed out of those of $SO(2N+1)$,  that transform them appropriately (Appendix).   Secondly,  the flux matching argument of Section~\ref{matching}  do connect these vortices to the minimum, regular monopoles appearing in the semiclassical analysis.   
As in the $D_{N}$ theories these observations should be considered at best as a modest hint that dual group structure as suggested by the monopole-vortex correspondence is consistent with the GNO conjecture.

\section{Conclusions}

In this paper we have explicitly constructed BPS, non-Abelian vortices
of a class of $SO(N)\times U(1)$  gauge theories in the Higgs phase. The models considered here can be regarded as the bosonic part of softly broken ${\cal N}=2$ gauge theories with  $N_{f}$ quark matter fields.
The vortices considered here represent non-trivial generalizations of the non-Abelian vortices in $U(N)$ models widely studied in recent literature.    

The systems are constructed so that they arise as  low-energy approximations to theories in which gauge symmetry suffers from a hierarchical breaking 
\begin{equation} SO(N+2) \stackrel{v_1}{\longrightarrow} SO(N) \times
  U(1) \stackrel{v_2}{\longrightarrow} {\mathbbm {1}}\ , \qquad v_{1}
  \gg  v_{2}\ ,   \label{hierarchylast} \end{equation} 
leaving an exact, unbroken  global $ (SO(N)\times U(1))_{C+F}$  symmetry.    Even though the low-energy $SO(N)\times U(1)$  model with symmetry breaking 
\begin{equation}  SO(N) \times U(1) \stackrel{v_2}{\longrightarrow}
  {\mathbbm {1}}\ ,   \label{nohierarchy} \end{equation} 
can be studied on its own right, without ever referring  to the high-energy $SO(N+2)$  theory,  consideration 
of the system with hierarchical symmetry breaking is interesting as it forces us to try  (and hopefully allows us)  to understand the properties of the non-Abelian {\it monopoles}  in the high-energy approximate system
with $SO(N+2) \stackrel{v_1}{\longrightarrow} SO(N) \times U(1)$ 
and their confinement by the vortices  -- language adequate in the dual variables --  from the properties of the vortices via homotopy map and symmetry argument.  Note that in this argument, the fact that the monopoles in the high-energy theory and the vortices in the low-energy theory  are both almost BPS but not exactly so, is of fundamental importance \cite{Duality,Strassler}.  


In the models based on $SU(N)$  gauge symmetry, the efforts along this line of thought  seem to be starting to give fruits,  giving
 some hints on the nature of non-Abelian duality and confinement. 
Although the results of this paper are a only a small step toward a better and systematic understanding of these 
questions  in a more general class of gauge systems, they provide a concrete starting point for further studies.  


\section*{Acknowledgement} 

This work is based on a master thesis by one of us (L.F.) \cite{LF}. The authors acknowledge useful discussions with Minoru Eto, Muneto Nitta, Giampiero Paffuti  and 
Walter Vinci. L.F thanks also Roberto Auzzi, Stefano Bolognesi, Jarah Evslin and Giacomo Marmorini for useful discussions and advices.


\appendix

\section{$SO(2N)$, $ USp(2N)$, $SO(2N+1)$   \label{grouptheory}}

The change of basis to the one  where  a vector  multiplet  $\underline{\mathbf{2N}}$  of $SO(2N)$ 
naturally breaks to $\underline{\mathbf{N}}+\underline{\mathbf{\bar{N}}}$ under $U(N)$, is given by   (see Eq.~(\ref{U2Ntra}))
\beq   \left(\begin{array}{c}{\hat Q}_3 \\\vdots \\{\hat Q}_{2N+1}          \\-i  Q_{3}  \\\vdots \\-i Q_{2N+1}\end{array}\right)  =
          \left( \begin{array}{cc}
                  {\mathbbm {1}}  /\sqrt{2} & -i    {\mathbbm {1}} /\sqrt{2}\\
                  -i    {\mathbbm {1}} /\sqrt{2} &    {\mathbbm {1}} /\sqrt{2}
          \end{array}
          \right)
          \, \left(\begin{array}{c}q_3  \\\vdots \\q_{2N+1} \\q_4
          \\\vdots \\q_{2N+2}\end{array}\right) \ .  \label{relatedby}
\eeq
 The $SO(2N)$ generators, 
\begin{equation}
          \left( \begin{array}{cc}
                  E & F \\ -\,^{t}\!F & D
          \end{array}
          \right),
\end{equation}
where $D$, $E$, $F$ are all pure imaginary $N \times N$ matrices, with
the constraints $ ^{t}\!E = -E$, $ ^{t}\!D = -D$, are accordingly transformed as 
\begin{eqnarray}
          \lefteqn{
          \left( \begin{array}{cc}
                  1/\sqrt{2} & -i/\sqrt{2}\\
                  -i/\sqrt{2} & 1/\sqrt{2}
          \end{array}
          \right)
          \left( \begin{array}{cc}
                  E & F \\ -\, ^{t}\!F & D
          \end{array}
          \right)
          \left( \begin{array}{cc}
                  1/\sqrt{2} & i/\sqrt{2}\\
                  i/\sqrt{2} & 1/\sqrt{2}
          \end{array}
          \right) } \nonumber \\
          & &
          = \frac{1}{2}
          \left( \begin{array}{cc}
                  (E+D) + i(F+\,^{t}\!F) & i(E-D)+(F-\,^{t}\!F)\\
                  -i(E-D)+(F-\,^{t}\!F) & (E+D) -i(F+\, ^{t}\!F)
          \end{array}
          \right) .  \label{ndbloks}
\end{eqnarray}
Since both $E$, $D$ are anti-symmetric, $(E+D)$ in the 1st block is
the most general anti-symmetric imaginary matrix, while $i(F+\, ^{t}\!F)$
is the most general symmetric real matrix.
Their sum gives the most general $N \times N$   hermitian matrix, which corresponds to   generators of $U(N)$.  In other words, the subgroup 
 $U(N) \subset SO(2N)$  is generated by those elements with   $E=D$, $F= \, ^{t}\!F$.

  On the other hand, the generators of
$USp(2N)$  group 
have the form
\begin{equation}
          \left( \begin{array}{cc}
                  B & A \\ C & -\,^{t}\!B
          \end{array}
          \right),  \label{theform}
\end{equation}
with the constraints, $^{t}\!A=A$, $^{t} C=C$, $A^{*}=C$,
$B^{\dagger}=B$.   The fact that  $A$ is symmetric while the non-diagonal blocks in Eq.~(\ref{ndbloks}) are antisymmetric, means that 
there is no further overlap between the two groups,  that is,  the maximal common subgroup between $SO(2N)$ and $USp(2N)$ is $U(N)$.

It is possible to get a hint on how $USp(2N)$ groups can appear as transformation group of the 
vortices. In order to see transformations among the vortices  (${\hat Q}, Q$) under which the latter could transform as $\underline{\mathbf{2N}}$, it is necessary to embed the system in  a larger group, such as  $SO(2N+1)$ model  considered in Section~\ref{Modd}.  
 The idea is to build a map \footnote{This correspondence can be applied equally well to the minimal regular monopoles constructed semi-classically, and 
has been  discussed in this context in \cite{FK}.} between the $SO(2N+1)$ generators (antisymmetric matrices) and the $USp(2N)$ generators which have the form, Eq.~(\ref{theform}).   
 The  $i$th   $SO(4)\sim SU(2)\times SU(2)$ subgroup is  generated by
(with a simplified notation  $(1,2,3,4) \equiv (1,2,2i+1,2i+2)$)
\begin{equation}    T_{1}^{\pm} = - \frac{i}{2}  \, (\Sigma_{2 3} \pm
 \Sigma_{ 4 1})\ , \quad T_{2}^{\pm} = - \frac{i}{2}  \, (\Sigma_{ 3
 1} \pm \Sigma_{ 4 2})\ , \quad T_{3}^{\pm} = - \frac{i}{2}  \,
 (\Sigma_{12} \pm \Sigma_{4 3}) \ . 
\end{equation} 
The two vortices 
 living in   this $SO(4)$ group are taken to be  $i$-th and ($N+i$)-th components of  the fundamental representation of 
$USp(2N)$. The pairs can be transformed to each other by rotations  in the $(2i+2, 2N+3)$
plane ($\subset SO(2N+1)$), thus
\begin{equation}       A_{i, i} = -i \, \Sigma_{2i+2, 2N+3} \ . 
\label{map1}\end{equation} 
On the other hand,  the two vortices  associated with subgroups $T^{\pm}$ living in the $(1,2,2i+1,2i+2)$ subspace and those living in the 
$(1,2,2j+1,2j+2)$ subspace, $j\ne i$, are transformed into each other  by rotations  in the
 $(2i+1, 2i+2, 2j+1, 2j+2)$ space: they transform in $SO(2N)$ (in the subspace $i=3,4,\ldots, 2N+2$).  We have already seen that they actually do transform as a pair of 
$U(N)$ representations,  in  the basis Eq.~(\ref{relatedby}). 
As  the $U(N)$ elements are generated by the $SO(2N)$ 
infinitesimal transformations with  $E=D$, $F= \,^{t}\!F$,    one finds the map,   
\begin{equation} B_{i, j} =     -i \,   (\Sigma_{2i, 2j}
  +\Sigma_{2i+1, 2j+1}) +  ( \Sigma_{2i, 2j+1} - \Sigma_{2i+1, 2j})\ . \label{map2}
\end{equation} 
Non-diagonal elements $A_{ij}$, $i \ne j$,  can be generated by commuting the actions of  (\ref{map1}) and (\ref{map2}).

\section{Fundamental groups}\label{homotopy}

Let's briefly discuss
the (first) homotopy groups  relevant to us: 

\subsection {$SO(2N+2)$\label{homoprimo}}   
 There is  only one non-trivial closed path $P$ in this case, the
 rotation from 0 to $2\pi$ around any axis. The rotation from 0 to
 $4\pi$ is homotopically equivalent to the trivial path, so
 $P^2=\mathbbm{1}$ and the homotopy group is
 \be{\pi_1\left(SO(2N+2)\right)={\mathbb {Z}}_2 \ , }  

\subsection { $SO(2N+2)/{\mathbb {Z}}_2$  \label{homso2np2}} 
Actually, in the model discussed in this paper, all the fields are in the adjoint representation of $SO(2N+2)$:   
 the gauge group effectively corresponds to $SO(2N+2)$ modulo identification $-\mathbbm{1}=\mathbbm{1}$. The path $P$ is again non-trivial, but now there are also two inequivalent closed paths $P_+$ and $P_-$ going from $\mathbbm{1}$ to  $-\mathbbm{1}$, defined as $P_+ P_-^{-1}=P$.   Explicitly,  they can be taken as  simultaneous rotations in $N+1$  planes 
\begin{align}
P_{+}:&  \quad  e^{i \beta_{12} \Sigma_{12}}\prod_{i=3,5,
 \ldots, N-1}   e^{i \beta_{i \, i+1} \Sigma_{i, i+1}}\ ; \qquad
 \beta_{12}: 0 \to \pi\ , \phantom{-}\quad  \beta_{i, i+1}: 0 \to \pi\ . \\
P_{-}:&  \quad  e^{i \beta_{12} \Sigma_{12}} \prod_{i=3,5,
 \ldots,  N-1}  e^{i \beta_{i \, i+1} \Sigma_{i, i+1}}\ ; \qquad
 \beta_{12}: 0 \to -\pi\ , \quad  \beta_{i, i+1}: 0 \to \pi\ .
\end{align}
When $N+1$ is even, $P_+^2=P_-^2=\mathbbm{1}$ and $P_+P_-=P$.  The homotopy group is generated by $P_+, P_{-}$: 
\be{\pi_1\left(\frac{SO(4N)}{{\mathbb {Z}}_2}\right)={\mathbb
 {Z}}_2\times {\mathbb {Z}}_2\ , }
When $N+1$ is odd, $P_+^2=P_-^2=P$ and $P_+P_-=\mathbbm{1}$, so the homotopy group is generated by $P_{+}$ only, and is of cyclic order
four 
\be{\pi_1\left(\frac{SO(4N+2)}{{\mathbb {Z}}_2}\right)={\mathbb
 {Z}}_4\ .}

\subsection {$(SO(2N)\times U(1))/{\mathbb {Z}}_2$}  
After the symmetry breaking at the higher mass scale $v_{1}$,  the theory reduces to an $(SO(2N)\times U(1))/{\mathbb {Z}}_2$
theory.  The division by ${\mathbb {Z}}_{2}$ corresponds to  the identification $(-\mathbbm{1},-1)=(\mathbbm{1},1)$, 
inherited from the underlying theory.  From the point of view of  the low-energy
effective theory,  it is due to the fact that all the light matter fields $q_{A, j}, {\tilde q}_{A,j}$ are in the vector representation of 
 $SO(2N)$ but they carry at the same time the unit charge with respect to $U(1)$.

   The non-trivial paths of $SO(2N)\times U(1)$ are combinations of $Q$   (a $2 \pi$ rotation in any  plane  in  $SO(2N)$)   and the paths $R_n$ winding $n$ times around the $U(1)$.   The simplest non-trivial closed paths that arise after the ${\mathbb {Z}}_2$ quotient are $P_{+, \frac{1}{2}}$, $P_{+,- \frac{1}{2}}$, $P_{-,\frac{1}{2}}$, $P_{-,- \frac{1}{2}}$ going from $(\mathbbm{1},1)$ to $(-\mathbbm{1},-1)$ with  
a half winding around $U(1)$.   By taking   $U(1)$ to act in the $(12)$  plane,  $SO(2N)$ in the $(34 \ldots N)$ space,   they can be explicitly  chosen as simultaneous rotations in $(12)$, $(34)$, $(56) \ldots$ planes 
\beq    \quad  e^{i \gamma_{12} \Sigma_{12}} \, e^{i \beta_{34}
  \Sigma_{34}} \!\prod_{i=5,7, \ldots, N-1} \!  e^{i \beta_{i \, i+1}
  \Sigma_{i, i+1}}\ ; 
 \eeq
with 
\begin{align}
P_{+,  \frac{1}{2}}:&   \quad \gamma_{12}: 0 \to \pi\ , \phantom{-}\quad
 \beta_{3 4}: 0 \to \pi\ ,\phantom{-}\quad \beta_{i, i+1}: 0 \to \pi\ .\\
P_{+,  -\frac{1}{2}}:&    \quad \gamma_{12}: 0 \to -\pi\ ,
 \quad  \beta_{3 4}: 0 \to \pi\ ,\phantom{-}\quad \beta_{i, i+1}: 0 \to \pi\ .\\
P_{-,  \frac{1}{2}}:&   \quad \gamma_{12}: 0 \to \pi\ , \phantom{-}\quad
 \beta_{3 4}: 0 \to -\pi\ ,  \quad \beta_{i, i+1}: 0 \to \pi\ .\\
P_{-,  -\frac{1}{2}}:&   \quad \gamma_{12}: 0 \to -\pi\ , \quad
 \beta_{3 4}: 0 \to -\pi\ ,  \quad \beta_{i, i+1}: 0 \to \pi\ .
\end{align}
 Note that $ P_{+,  \frac{1}{2}}$ and $ P_{+,  -\frac{1}{2}}$
 correspond respectively to the $P_{+}$ and $P_{-}$ paths in
 the $SO(2N+2)$  theory.  
 
  When $N$ is even, $P_{+,a}P_{+,b}=P_{-,a}P_{-,b}=R_{a+b}$ and $P_{+,a}P_{-,b}=Q\, R_{a+b}$, so every group element can  be written as $(P_{+,1/2})^k \, Q^\delta$ with $k \in \mathbb{Z}$, $\delta=\{0,1\}$.
The homotopy group is 
\be{\pi_1\left(\frac{SO(2N)\times U(1)}{{\mathbb
      {Z}}_2}\right)=\mathbbm{Z}\times {\mathbb {Z}}_2 \ ,\quad N
  \mbox{ even}\ , }
When $N$ is odd, $P_{+,a}P_{+,b}=P_{-,a}P_{-,b}=Q\, R_{a+b}$ and
$P_{+,a}P_{-,b}=R_{a+b}$, and every group element can again be written
as $(P_{+,1/2})^k \, Q^\delta$ with $k \in \mathbb{Z}$, $\delta=\{0,1\}$,   as
in the $N$ even case.
The homotopy group is 
\be{\pi_1\left(\frac{SO(2N)\times U(1)}{{\mathbb
      {Z}}_2}\right)=\mathbbm{Z}\times {\mathbb {Z}}_2 \quad,\quad N
  \mbox{ odd}\ , }
Even though the homotopy group is the same for the two cases ($N$ even or odd), its embedding in $\pi_1\left(SO(2N)\times U(1)\right)=\mathbbm{Z}\times {\mathbb {Z}}_2$ is different: $R_n$ corresponds to $k=2n,\delta=0$ for $N$ even and to $k=2n,\delta=1$ for $N$ odd. In other words
\beq   R_{1}=   (P_{+,1/2})^2 \, Q\ , \quad (N \,{\rm odd})\ ; \qquad
R_{1}=   (P_{+,1/2})^2 \quad (N \,{\rm even})\ . \label{useful} 
\eeq

\subsection   {Relation between the smallest elements of the high-energy and low-energy fundamental groups} 

 There are simple relations among the smallest elements of the groups   $\pi_1\left(\frac{SO(2N+2)}{{\mathbb {Z}}_2}\right) $ and   $\pi_1\left(\frac{SO(2N)\times U(1)}{{\mathbb {Z}}_2}\right)$.  From the above explicit constructions one sees that 
\beq     P_{+} = P_{+,  \frac{1}{2}}\ ; \quad   P_{-}=  P_{+,
  -\frac{1}{2}}=  R_{-1} \,P_{+,  \frac{1}{2}}\ ; 
\eeq
and by using Eq.~(\ref{useful}),  one has 
\beq   P_{+,  -\frac{1}{2}}=  \begin{cases}
     (P_{+,  \frac{1}{2}})^{-1}\,  Q\ ,  & \text{odd $N$}\ , \\
    (P_{+,  \frac{1}{2}})^{-1}\ ,      & \text{even $N$}\ .
\end{cases}
\eeq

\subsection { $SO(2N+3)$}   The fundamental group is ${\mathbb {Z}}_{2}$ as in the $SO(2N+2)$ cases, and the smallest closed path being  
\beq   P:  \quad  e^{i \beta_{ij} \Sigma_{ij}}:  \quad   \beta_{ij} =
 0 \to 2 \pi\ ,
 \eeq
 in any plane $(ij)$.   $P^2=\mathbbm{1}$ and the homotopy group is
 \be{\pi_1\left(SO(2N+3)\right)={\mathbb {Z}}_2 \ .}

\subsection {$SO(2N+1)\times U(1)$}   
At the mass scales below $v_{1}$ the theory reduces to an $SO(2N+1)\times U(1)$ theory with matter in the fundamental representation,   $q $ and ${\tilde q} $  carrying charges $\pm 1$   with respect to $U(1)$.     The fundamental  group is 
\beq  \pi_{1}\left(SO(2N+1)\times U(1)\right)=  {\mathbb {Z}}_{2}\times {\mathbb
  {Z}}\ , 
\eeq 
where ${\mathbb {Z}}$ represents the number of winding (charge) in the $U(1)$ part and ${\mathbb {Z}}_{2}$ a $2\pi$ rotation in any 
plane in $SO(2N+1)$.  










\begin{thebibliography}{60} 


\bibitem{NAmonop}
  E. Lubkin,  { Ann. Phys.  {\bf  23}},  233  (1963);   E. Corrigan, D.I. Olive, D.B. Fairlie,  
  J. Nuyts,  Nucl. Phys. B  {\bf   106},  475 (1976)

\bibitem{GNO}   P. Goddard, J. Nuyts,  D. Olive,    Nucl. Phys. B {\bf  125}, 1 
(1977) 

\bibitem{BS}  F.A. Bais,   Phys. Rev. D {\bf  18},  1206  (1978)
 
\bibitem{EW}   E.J. Weinberg,  Nucl. Phys.  B {\bf  167}, 500 (1980);  Nucl. Phys. B {\bf 203}, 445  (1982); 
K. Lee, E. J. Weinberg, P. Yi,   Phys. Rev. D  {\bf  54 }, 6351  (1996)

\bibitem{Coleman}  S. Coleman, ``The Magnetic Monopole  Fifty Years Later'', 
Lectures given at Int. Sch. of Subnuclear Phys.,  Erice, Italy (1981)  

\bibitem{CDyons} A. Abouelsaood,  Nucl. Phys.  B {\bf 226},  309 (1983);  P. Nelson, A. Manohar,  Phys. Rev. Lett. {\bf  50},  943
(1983);  A. Balachandran, G. Marmo, M. Mukunda, J. Nilsson, E. Sudarshan, F. Zaccaria,    Phys. Rev. Lett. {\bf  50},  1553
(1983);  P. Nelson, S. Coleman,  Nucl. Phys.  B {\bf 227},  1  (1984)

\bibitem{DFHK} N.~Dorey, C.~Fraser, T.J.~Hollowood and M.A.C.~Kneipp,
 ``Non-Abelian duality in {\cal N}=4 supersymmetric gauge theories,''  arXiv:hep-th/9512116;   { Phys.Lett. B383}  (1996) 422 [arXiv:hep-th/9605069].

  \bibitem{ABEKM}
R.~ Auzzi, S.~Bolognesi, J.~Evslin, K.~Konishi and  H.~Murayama,  { Nucl. Phys. B701}  (2004) 207
[hep-th/0405070].

\bibitem{Duality} 
M. Eto, L. Ferretti, K. Konishi, G. Marmorini, M.  Nitta, K. Ohashi, W. Vinci,  N.  Yokoi,
``Non-Abelian duality from vortex moduli: A dual model of color-confinement'',
 [hep-th/0611313], Nucl. Phys. B (2007), to appear;  
 
 \bibitem{Konishi}   K. Konishi, ``Magnetic Monopole Seventy-Five Years Later'', 
to appear in a special volume of Lecture Notes in Physics, Springer, in honor of the 65th birthday of Gabriele Veneziano,  [hep-th/0702102]


\bibitem{APS}
P. C. Argyres, M. R. Plesser,  N. Seiberg, Nucl. Phys.  B {\bf 471}, 159  
(1996);   
P.C. Argyres, M.R. Plesser,  A.D. Shapere, 
Nucl. Phys. B {\bf 483}, 172 (1997);
 K.  Hori, H. Ooguri,   Y.  Oz,
 Adv. Theor. Math. Phys.   {\bf {1}}, 1  (1998)

\bibitem{HO}  A. Hanany,  Y. Oz, 
  Nucl. Phys.  B {\bf  466},  85  (1996)

\bibitem{CKM}
G. Carlino, K. Konishi,  H. Murayama,
     JHEP  {\bf  0002}, 004    (2000);  
 Nucl. Phys.   B {\bf 590},  37   (2000); 
G. Carlino, K.  Konishi, S. P.  Kumar, H.  Murayama,  
Nucl.  Phys.  B {\bf 608}, 51 (2001) 

\bibitem{BKM}   S. Bolognesi, K. Konishi,   Nucl. Phys. B {\bf  645},  337 (2002), 
S. Bolognesi, K. Konishi,   G. Marmorini,  
Nucl. Phys. B  {\bf 718}, 134  (2005)


\bibitem{MKY}
 G. Marmorini, K. Konishi,  N. Yokoi,  Nucl. Phys.  B {\bf 741},  180 (2006) 
 
  \bibitem{Hashimoto}
  K.~Hashimoto and D.~Tong,
    { JCAP  0509} (2005)  004
  [arXiv:hep-th/0506022].

\bibitem{Auzzi:2005gr}
R.~Auzzi, M.~Shifman and A.~Yung,  { Phys. Rev. D73} (2006) 105012
  [arXiv:hep-th/0511150].
  
  \bibitem{seven}
 M. Eto, K. Konishi, G. Marmorini, M. Nitta, K. Ohashi, W. Vinci,  N. Yokoi,
   Phys. Rev.  D {\bf  74}, 065021 (2006) 
     [arXiv:hep-th/0607070].

 
 
 \bibitem{HT}
  A. Hanany,  D. Tong,
  JHEP {\bf  0307}, 037 (2003); 
A. Hanany,  D. Tong, JHEP {\bf 0404}, 066  (2004)  

 
 \bibitem{ABEKY}
  R. Auzzi, S. Bolognesi, J. Evslin, K. Konishi,  A. Yung,
  Nucl. Phys. B {\bf  673}, 187 (2003)
  
  \bibitem{SY}  
  M. Shifman and A. Yung,  Phys. Rev.  D {\bf 70},  045004  (2004)
A. Gorsky,  M. Shifman,   A. Yung,  Phys. Rev. D  {\bf 71},    045010 (2005)

\bibitem{ABEK}
R.~Auzzi, S.~Bolognesi, J.~Evslin and K.~Konishi,
{ Nucl. Phys. B686}  (2004)  119  
[hep-th/0312233].

\bibitem{Tong} D. Tong,
  ``TASI lectures on solitons: Instantons, monopoles, vortices and kinks''
[arXiv: hep-th/0509216.


   \bibitem{Isozumi:2004vg}
  Y. Isozumi, M. Nitta, K. Ohashi,  N. Sakai,
   Phys. Rev. D {\bf 71}, 065018  (2005) 

     \bibitem{FK}
  L.~Ferretti and   K.~Konishi,
 ``Duality and confinement in SO(N) gauge theories'',
 FestSchrift, ``Sense of Beauty in Physics,'' in honor of the 70th birthday of A. Di Giacomo,
  Edizioni PLUS (University of Pisa Press), 2006  
  [arXiv: hep-th/0602252].
  
      \bibitem{LF} Luca Ferretti,
``Vortici non abeliani e gruppi duali in teorie di gauge $N=2$ $SO(N)$ e $USp(2N)$'',  
master thesis at University of Pisa, 2004  (unpublished).
  

\bibitem{Georgi}
H.~Georgi,  ``Lie Algebras in Particle Physics'', Second Ed., Westview Press (1999). 



\bibitem{Strassler}
M. J.~Strassler,  JHEP { 9809}  (1998)  017   
[arXiv:hep-th/9709081];
``On Phases of Gauge Theories and the Role of Non-BPS Solitons in Field Theory '',
III Workshop, ``Continuous Advance in QCD'', Univ. of Minnesota  (1998)  [arXiv: 
[arXiv:hep-th/9808073].


\bibitem{Eto:2006db}
  M.~Eto, K.~Hashimoto, G.~Marmorini, M.~Nitta, K.~Ohashi and W.~Vinci,
  Phys.\ Rev.\ Lett.\  {\bf 98} (2007) 091602
  [arXiv:hep-th/0609214].


  

  

  

%
  
\end{thebibliography}
\end{document}